# Grover's algorithm on a Feynman computer.

Diego de Falco

Dario Tamascelli

Dipartimento di Scienze dell' Informazione
Università di Milano
Via Comelico 39
20135 Milano, Italy

**Abstract**: We present an implementation of Grover's algorithm in the framework of Feynman's cursor model of a quantum computer. The cursor degrees of freedom act as a quantum clocking mechanism, and allow Grover's algorithm to be performed using a single, time-independent Hamiltonian. We examine issues of locality and resource usage in implementing such a Hamiltonian. In the familiar language of Heisenberg spin-spin coupling, the clocking mechanism appears as an excitation of a basically linear chain of spins, with occasional controlled jumps that allow for motion on a planar graph: in this sense our model implements the idea of "timing" a quantum algorithm using a continuous-time random walk. In this context we examine some consequences of the entanglement between the states of the input/output register and the states of the quantum clock.

## 1. Introduction.

The starting point of our discussion is the analysis of the physical aspects of Grover's algorithm given in References [1] and [2].

Suppose one is given an "oracle" able to compute, in a quantum reversible way, the indicator function of a binary word $\mathbf{a}$ of an assigned length $\mu$. We will assume, for the sake of definiteness, that this computation is performed by applying a unitary transformation $\mathsf{A}$ to the states of a "register" consisting of spin 1/2 systems.

Suppose that $\mathsf{A}$ results from the action, for a fixed amount $\bar{t}$ of time, of a Hamiltonian $K(\mathbf{a})$:

$$\mathsf{A} = \exp(-i \cdot \bar{t} \cdot K(\mathbf{a})). \qquad (1.1)$$

It is then possible to arrange things in such a way that the state $|\mathbf{a}\rangle$ that corresponds to having "the word $\mathbf{a}$ written on the register", is the ground state of $K(\mathbf{a})$.

The search for the ground state of $K(\mathbf{a})$ is performed, in Reference [2], following the simple idea of perturbing the Hamiltonian $K(\mathbf{a})$,

$$K(\mathbf{a}) \to K(\mathbf{a}) + \beta \ , \qquad (1.2)$$

with a perturbation $\beta$ chosen in such a way that a suitable initial condition <u>oscillates</u> about the state $|\mathbf{a}\rangle$ with a period proportional to $2^{\mu/2}$, becoming, at a time $O(2^{\mu/2})$, parallel to the target state.

By applying Trotter's product formula to $\exp(-i \cdot t \cdot (K(\mathbf{a}) + \beta))$, it is shown, in Reference [1], that no significant loss in the probability $\Pr(t)$ of finding $\mathbf{a}$, at suitable values of time $t$, results from alternating intervals of time in which only the "oracle" Hamiltonian $K(\mathbf{a})$ is active, thus in fact applying the "oracle" transformation $\mathsf{A} = \exp(-i \cdot \bar{t} \cdot K(\mathbf{a}))$, with intervals in which only $\beta$ is active, applying, in fact, the "estimator" transformation

$$\mathsf{B} = \exp(-i \cdot \bar{t} \cdot \beta). \qquad (1.3)$$

The <u>oscillatory</u> nature of the quantum search algorithm is confirmed, in this discrete time setting, by the analysis of Reference [3].

In this paper we examine some simple models, variants of Feynman's cursor model [4], in which the physical agent that alternatively administers the transformations $\mathsf{A}$ and $\mathsf{B}$ to the quantum register is itself a quantum system.

The models of quantum control mechanisms that we consider here are highly idealised (as compared, say, to the quantized electromagnetic field modes used in [6] to explore some aspects of Grover's algorithm). They are, however, simple enough to allow for explicit expressions, as a function of time, of the probability $\Pr(t)$ of finding $\mathbf{a}$ on the register.

For the models that we are going to consider, we will show, using techniques developed in [7], that the overlap probability $\Pr(t)$ between the state at time $t$ and the target state admits, in a suitable time scale, asymptotic expressions of the form

$$\Pr(t) \cong \frac{1}{2} - \frac{1}{2} \cdot \left( J_0\left(\frac{t}{2^{\mu/2}}\right) - J_2\left(\frac{t}{2^{\mu/2}}\right) \right) \tag{1.4}$$

where $J_0$ and $J_2$ are Bessel functions of the first kind.

A plot of 1.4 , given in figure 1, shows the effect of a quantum control mechanism that we will try to understand in this paper: there are large intervals of time over which the oscillations of the state of the register about the ground state appear to be <u>damped</u>.

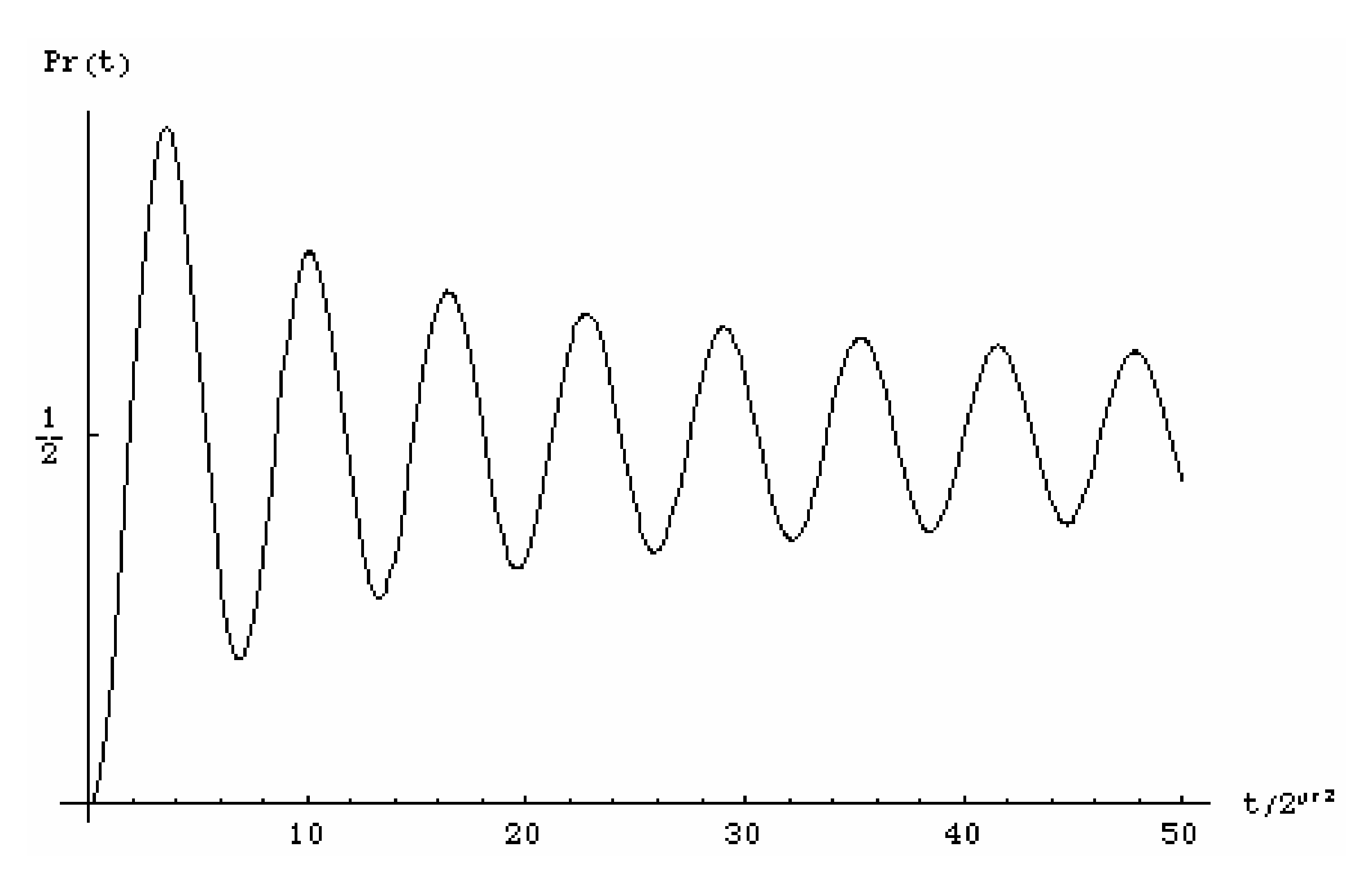

Figure 1

$$\Pr(t) \cong \frac{1}{2} - \frac{1}{2} \cdot \left( J_0\left(\frac{t}{2^{\mu/2}}\right) - J_2\left(\frac{t}{2^{\mu/2}}\right) \right)$$

The paper is organized as follows.

In section 2 we study the simplest model, a linear chain quantum walk [5] able to perform Grover's algorithm.

In section 3 we study a family of continuous-time quantum walks arising from Feynman's suggestion [4] of a quantum clocking mechanism able to implement, at the quantum level, the notion of "subroutine".

Grover's algorithm, with its alternance, for a prescribed large number of times, of the action of an "oracle" A and an "estimator" B, provides, indeed, the natural testing ground for quantum subroutines.

In sections 2 and 3 both the "oracle" A and the "estimator" B are treated as "primitives" that can be performed in a single "step" of the clocking mechanism. In section 4 we decompose them into more elementary "steps" involving only the two reversible primitives SWITCH (a reversible version of a conditional jump) and NOT; we examine the computational costs involved in this decomposition .

Section 5 is devoted to numerical examples, section 6 to concluding remarks.

## 2. The linear chain model.

Let $\mu$ be a positive integer. Set:

$$\nu = \mu + 1 \ . \tag{2.1}$$

The input/output register of all the models that we are going consider will be formed by a collection $\underline{\sigma}(1), \underline{\sigma}(2),..., \underline{\sigma}(\mu), \underline{\sigma}(\nu)$ of spin $1/2$ systems.

We denote by $(\sigma_1(i),\ \sigma_2(i),\ \sigma_3(i)) \equiv (\sigma_x(i),\ \sigma_y(i),\ \sigma_z(i))$ the three components of $\underline{\sigma}(i)$ with respect to an assigned reference frame, and by $H_{register}$ the $2^{\nu}$ dimensional state space of the "register degrees of freedom".

Let $\mathbf{a} = (a_1, a_2, \ldots, a_\mu) \in \{-1,1\}^{\mu}$ be a fixed binary word of length $\mu$.

With reference to the fixed word $\mathbf{a}$, define a linear operator A: $H_{register} \to H_{register}$ through its action on the simultaneous eigenstates of $(\sigma_3(1),\ \sigma_3(2),\ldots,\ \sigma_3(\mu),\ \sigma_3(\nu))$:

$$\begin{aligned} &\mathrm{A}\left|\sigma_3(1) = z_1, \ldots, \sigma_3(\mu) = z_\mu, \sigma_3(\nu) = z_\nu\right\rangle = \\ &= \begin{cases} \left|\sigma_3(1) = z_1, \ldots, \sigma_3(\mu) = z_\mu, \sigma_3(\nu) = -z_\nu\right\rangle \text{ if } \mathbf{z} = \mathbf{a} \\ \left|\sigma_3(1) = z_1, \ldots, \sigma_3(\mu) = z_\mu, \sigma_3(\nu) = \ z_\nu\right\rangle \text{ if } \mathbf{z} \neq \mathbf{a} \end{cases} \end{aligned} \tag{2.2}$$

where $\mathbf{z} = (z_1, z_2, \ldots, z_\mu) \in \{-1,1\}^{\mu}$, $z_\nu \in \{-1,1\}$.

The "oracle" A performs a quantum reversible computation of the indicator function of $\mathbf{a}$ by flipping the $z$ component of the output qubit $\underline{\sigma}(\nu)$ iff the word $\mathbf{a}$ is written on the register in terms of the $z$ components of the input qubits $\underline{\sigma}(1), \underline{\sigma}(2),..., \underline{\sigma}(\mu)$.

Define, in a similar way, a linear operator B: $H_{register} \rightarrow H_{register}$ through the following action on the simultaneous eigenstates of $\left(\sigma_1(1),\ \sigma_1(2),\ldots,\ \sigma_1(\mu),\ \sigma_3(\nu)\right)$:

$$\mathrm{B}\left|\sigma_1(1)=x_1,\ldots,\sigma_1(\mu)=x_\mu,\sigma_3(\nu)=z_\nu\right\rangle = \\ =\begin{cases}\left|\sigma_1(1)=x_1,\ldots,\sigma_1(\mu)=x_\mu,\sigma_3(\nu)=-z_\nu\right\rangle \text{ if } \mathbf{x}=\mathbf{1}_\mu \\ \left|\sigma_1(1)=x_1,\ldots,\sigma_1(\mu)=x_\mu,\sigma_3(\nu)=\ z_\nu\right\rangle \text{ if } \mathbf{x}\neq\mathbf{1}_\mu\end{cases} \tag{2.3}$$

where $\mathbf{x}=\left(x_1,x_2,\ldots,x_\mu\right)\in\{-1,1\}^\mu$ , $z_\nu\in\{-1,1\}$ and $\mathbf{1}_\mu=(\underbrace{1,1,\ldots,1}_{\mu\ times})$.

The "estimator" B performs a quantum reversible computation of the indicator function of the word $\mathbf{1}_\mu$ by flipping the $z$ component of the output qubit $\underline{\sigma}(\nu)$ iff the word $\mathbf{1}_\mu$ is written on the register in terms of the $x$ components of the input qubits $\underline{\sigma}(1)$, $\underline{\sigma}(2)$,..., $\underline{\sigma}(\mu)$.

The quantum clocking mechanism that administers, in the correct order, the various primitives to the register is, in Feynman's cursor model [4], formed by a collection of a certain number $s$ of spin $1/2$ systems $\underline{\tau}(1)$, $\underline{\tau}(2)$,..., $\underline{\tau}(s)$, where

$$\underline{\tau}(j)=\left(\tau_1(j),\ \tau_2(j),\ \tau_3(j)\right)\equiv\left(\tau_x(j),\ \tau_y(j),\ \tau_z(j)\right) \quad j=1,\ldots,s. \tag{2.4}$$

We shall denote by $\tau_\pm(j)=\left(\tau_x(j)\pm i\cdot\tau_y(j)\right)/2$ the raising and lowering operators for the $z$ components of the cursor spins and by $H_{cursor}$ the $2^s$ dimensional state space of the "cursor degrees of freedom".

For each $j\in\{1,\ldots,s-1\}$ define a linear operator $A(j): H_{register}\otimes H_{cursor}\rightarrow H_{register}\otimes H_{cursor}$ through the position

$$A(j)=A_{forward}(j)+A_{backward}(j), \tag{2.5}$$

where

$$A_{forward}(j)=\mathrm{A}\otimes\tau_+(j+1)\cdot\tau_-(j) \tag{2.6}$$

and

$$A_{backward}(j)=\mathrm{A}^*\otimes\tau_+(j)\cdot\tau_-(j+1) \tag{2.7}$$

For each $j\in\{1,\ldots,s-1\}$ define a linear operator $B(j): H_{register}\otimes H_{cursor}\rightarrow H_{register}\otimes H_{cursor}$

through the analogous position

$$B(j) = B_{forward}(j) + B_{backward}(j) = \mathsf{B}\cdot \tau_{+}(j+1)\cdot \tau_{-}(j) + \mathsf{B}^{*}\cdot \tau_{+}(j)\cdot \tau_{-}(j+1) \qquad (2.8)$$

Set, furthermore,

$$C(j) = C_{forward}(j) + C_{backward}(j) \qquad (2.9)$$

where

$$C_{forward}(j) = \begin{cases} A_{forward}(j) & \text{if } j \text{ is odd} \\ B_{forward}(j) & \text{if } j \text{ is even} \end{cases} \quad \text{and} \quad C_{backward}(j) = \begin{cases} A_{backward}(j) & \text{if } j \text{ is odd} \\ B_{backward}(j) & \text{if } j \text{ is even} \end{cases} \qquad (2.10)$$

In this section we will study the evolution of the system formed by the register spins and by the cursor spins under the action of a Hamiltonian operator of the form

$$H = -\frac{\lambda}{2}\cdot\sum_{j=1}^{s-1} C(j), \qquad (2.11)$$

where, for notational convenience, we have indicated by $-\lambda/2$ a coupling constant.

A graphical representation of the Hamiltonian $H$ is given, for an odd value of $s$, in figure 2.

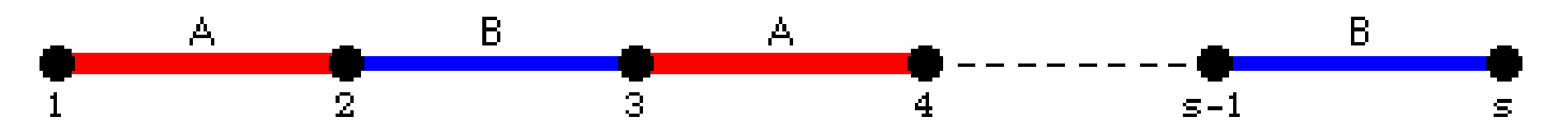

Figure 2
*The Hamiltonian 2.11 describes an $XY$ interaction between nearest neighbour cursor spins.*
$\mathsf{A}$ *is the "coupling constant" between spins corresponding to odd links;*
$\mathsf{B}$ *is the "coupling constant" between spins corresponding to even links.*
*Both* $\mathsf{A}$ *and* $\mathsf{B}$ *are, in fact, functions of the register spins.*

The solution of the Schrödinger equation

$$i\cdot\frac{d}{dt}|\,\psi(t)\rangle = H|\,\psi(t)\rangle \qquad (2.12)$$

under the initial condition

$$|\,\psi(0)\rangle = |\,\sigma_1(1)=1,\dots,\sigma_1(\mu)=1, \sigma_1(\nu)=-1\rangle \otimes |\,\tau_3(1)=1,\ \tau_3(2)=-1..,\tau_3(s)=-1\rangle \qquad (2.13)$$

is quite elementary because of the following conservation laws:

$$[H, \sigma_1(\nu)] = 0 \tag{2.14}$$

$$\left[H, \sum_{j=1}^{s} \tau_3(j)\right] = 0 \tag{2.15}$$

$$[H, P] = 0\,. \tag{2.16}$$

The operator $P$ in 2.16 is the projector $\sum_{j=1}^{s} |\varphi_j\rangle\langle\varphi_j|$ on the subspace spanned by the initial condition

$$|\varphi_1\rangle \equiv |\psi(0)\rangle \tag{2.17}$$

and by it logical successors $|\varphi_k\rangle$, which are defined [8], for $k = 2,\ldots,s$, by

$$|\varphi_k\rangle = \sum_{j=1}^{s-1} C_{forward}(j)|\varphi_{k-1}\rangle \;=\; C_{forward}(k)|\varphi_{k-1}\rangle \tag{2.18}$$

The solution of 2.12 under 2.13, will be, because of 2.16, of the form

$$|\psi(t)\rangle = \sum_{j=1}^{s} c(t,j;s)|\varphi_j\rangle \tag{2.19}$$

where the coefficients $c(t,\, j;\; \mathrm{s})$ satisfy the differential equations:

$$i\frac{d}{dt}c(t,j;s) = -\frac{\lambda}{2}\cdot\left(c(t,j-1;s) + c(t,j+1;s)\right) \quad \text{for} \quad 1 \le j \le s \tag{2.20}$$

under the boundary conditions

$$c(t,0;s) = 0 \tag{2.21}$$

$$c(t,s+1;s) = 0$$

(2.22)

and the initial condition

$$c(0,j;s) = \delta_{1,j} \quad \text{for} \quad 1 \le j \le s\,. \tag{2.23}$$

The explicit solution of equations 2.20-23 is [9]:

$$c(t,j;s) = \frac{2}{s+1}\sum_{n=1}^{s}\exp\left[i\cdot\lambda\cdot t\cdot\cos(\vartheta(n;s))\right]\cdot sin(\vartheta(n;s))\cdot sin(j\cdot\vartheta(n;s)) \qquad (2.24)$$

where

$$\vartheta(n;s) = \frac{n\cdot\pi}{s+1}$$

(2.25)

As the initial condition $|\varphi_1\rangle$ is an eigenstate, belonging to the eigenvalue 1, of the operator

$$Q = \sum_{j=1}^{s} j\cdot\frac{1+\tau_3(j)}{2}\ , \qquad (2.26)$$

its logical successor $|\varphi_k\rangle$ will be an eigenstate $|Q = k\rangle$ of $Q$ belonging to the eigenvalue $k$.

Because of 2.14 and 2.18, it is, then

$$|\varphi_k\rangle = \mathsf{C}(k-1)\ldots\mathsf{C}(2)\,\mathsf{C}(1)|\sigma_1(1)=1,\ldots,\sigma_1(\mu)=1\rangle\otimes|\sigma_1(\nu)=-1\rangle\otimes|Q=k\rangle \qquad (2.27)$$

where

$$\mathsf{C}(j) = \begin{cases}\mathsf{A} & if\ j\ is\ odd\\ \\ \mathsf{B} & if\ j\ is\ even.\end{cases} \qquad (2.28)$$

It helps, we think, to read the solution 2.19 in the following terms: an excitation (a single spin "up") of the linear chain $\underline{\tau}(1)$, $\underline{\tau}(2)$,..., $\underline{\tau}(s)$, for which the operator $Q$ defined in 2.26 has the meaning of a position operator, performs a quantum walk, ruled by equation 2.20, on the <u>sites</u> of figure 2; a transition of one step to the right has, according to 2.20, the same probability amplitude per unit time as a transition of one step to the left; because of the conservation law 2.16, either transition is accompanied by the application (to the state of the register spins $\underline{\sigma}(1)$, $\underline{\sigma}(2)$,..., $\underline{\sigma}(\mu)$ ) of the transformation associated to the <u>link</u> going from the initial site to the final site.

If $k$ is an odd number, $k = 2\cdot n+1$, it is $\mathsf{C}(k-1)\ldots\mathsf{C}(2)\,\mathsf{C}(1) = (\mathsf{B}\cdot\mathsf{A})^n$ .

An explicit expression for $(\mathsf{B}\cdot\mathsf{A})^n|\sigma_1(1)=1,\ldots,\sigma_1(\mu)=1\rangle$ can be found by the iterative procedure of Reference [3]:

$$(\mathsf{B}\cdot\mathsf{A})^n|\sigma_1(1)=1,\ldots,\sigma_1(\mu)=1\rangle = \left(\alpha_n(\mu)\cdot|\mathbf{a}\rangle_3 + \beta_n(\mu)\cdot\sum_{\mathbf{z}\neq\mathbf{a}}|\mathbf{z}\rangle_3\right), \qquad (2.29)$$

where

$$\alpha_n(\mu) = (-1)^n \cdot \sin((2 \cdot n + 1) \cdot \chi(\mu)) \tag{2.30}$$

$$\chi(\mu) = \arcsin\left(2^{-\mu/2}\right) \tag{2.31}$$

$$\beta_n(\mu) = \frac{(-1)^n}{\sqrt{2^\mu - 1}} \cdot \cos((2 \cdot n + 1) \cdot \chi(\mu)). \tag{2.32}$$

In 2.29 we have omitted explicit reference to $|\sigma_1(\nu) = -1\rangle$, as the conservation law 2.14 allows us to do, and we have set, for every $\mathbf{z} = (z_1, z_2, \ldots, z_\mu) \in \{-1,1\}^\mu$,

$$|\mathbf{z}\rangle_3 = |\sigma_3(1) = z_1, \ldots, \sigma_3(\mu) = z_\mu\rangle. \tag{2.33}$$

The case of an even value of $k$, $k = 2 \cdot n + 2$, can be analysed in a similar way, by observing that

$$\mathsf{A} \cdot (\mathsf{B} \cdot \mathsf{A})^n |\sigma_1(1) = 1, \ldots, \sigma_1(\mu) = 1\rangle = \left(-\alpha_n(\mu) \cdot |\mathbf{a}\rangle_3 + \beta_n(\mu) \cdot \sum_{\mathbf{z} \neq \mathbf{a}} |\mathbf{z}\rangle_3\right) \tag{2.34}$$

Summarizing, and considering, for the sake of definiteness, the case of an odd value of $s$, $s = 2 \cdot g + 1$, the solution of 2.12 under the initial condition 2.13 can be written as:

$$\begin{aligned} |\psi(t)\rangle &= \sum_{n=0}^{g} c(t, 2n+1; s)\left(\alpha_n(\mu)|\mathbf{a}\rangle_3 + \beta_n(\mu)\sum_{\mathbf{z} \neq \mathbf{a}}|\mathbf{z}\rangle_3\right) \otimes |\sigma_1(\nu) = -1\rangle \otimes |Q = 2n+1\rangle + \\ &+ \sum_{n=0}^{g-1} c(t, 2n+2; s) \cdot \left(-\alpha_n(\mu)|\mathbf{a}\rangle_3 + \beta_n(\mu)\sum_{\mathbf{z} \neq \mathbf{a}}|\mathbf{z}\rangle_3\right) \otimes |\sigma_1(\nu) = -1\rangle \otimes |Q = 2n+2\rangle \end{aligned} \tag{2.35}$$

where the amplitudes $c(t, j; s)$ are given by 2.23 and 2.24.

The probability $\Pr(t)$ that simultaneous measurements of $\sigma_3(1), \sigma_3(2), \ldots, \sigma_3(\mu)$ on the state $|\psi(t)\rangle$ give, respectively, the values $a_1, a_2, \ldots, a_\mu$ (namely the probability of finding the word $\mathbf{a}$ written, "in the $z$ direction", on the input part of the register) is therefore given by:

$$\begin{aligned} \Pr(t) &= \sum_{n=0}^{g} \alpha_n(\mu)^2 \cdot \left(|c(t, 2n+1; s)|^2 + |c(t, 2n+2; s)|^2\right) = \\ &= \sum_{x=1}^{s} |c(t, x; s)|^2 \cdot \left(\sin(\chi(\mu) \cdot x_{odd})\right)^2 \end{aligned} \tag{2.36}$$

If $x$ is a positive integer, we have denoted, in 2.36, by $x_{odd}$ the largest odd number not larger than $x$.

It is possible to give a simple approximation of 2.36 by means of the approximations for the coefficients $c(t,x;s)$ studied in Reference [7]. It is, indeed, for $s >> 1$ and for $0 < \lambda \cdot t < s$

$$|c(t,x;s)|^2 \approx \frac{4 \cdot x^2}{(\lambda \cdot t)^2} J_x(\lambda \cdot t)^2 \equiv f(t,x). \tag{2.37}$$

It is, furthermore, possible, for large values of $t$, to treat the discrete random variable having the probability mass function $f(t,x)$ as a continuous random variable having the probability density function

$$\rho(t,x) = \frac{4 \cdot x^2}{\pi \cdot (\lambda \cdot t)^2 \sqrt{(\lambda \cdot t)^2 - x^2}} I_{(0,\lambda \cdot t)}(x), \tag{2.38}$$

where $I_{(0,\lambda \cdot t)}(x)$ is the indicator function of the interval $(0,\ \lambda \cdot t)$.

In the context of these approximations,

$$\begin{aligned} \Pr(t) &\approx \int_0^{\lambda \cdot t} \frac{4 \cdot x^2}{\pi \cdot (\lambda \cdot t)^2 \sqrt{(\lambda \cdot t)^2 - x^2}} \cdot \left(sin(\chi(\mu) \cdot x)\right)^2 dx = \\ &= \frac{1}{2} - \frac{1}{2}\left(J_0(2 \cdot \chi(\mu) \cdot \lambda \cdot t) - J_2(2 \cdot \chi(\mu) \cdot \lambda \cdot t)\right) \end{aligned} \tag{2.39}$$

Figure 3 is a plot of the right hand side of 2.36 (the dashed graph) and of the right hand side of 2.39 (the solid line). With respect to the exact expression 2.36 for $\Pr(t)$, the approximate expression 2.39 misses the effects of the reflection taking place because of the boundary condition 2.22.

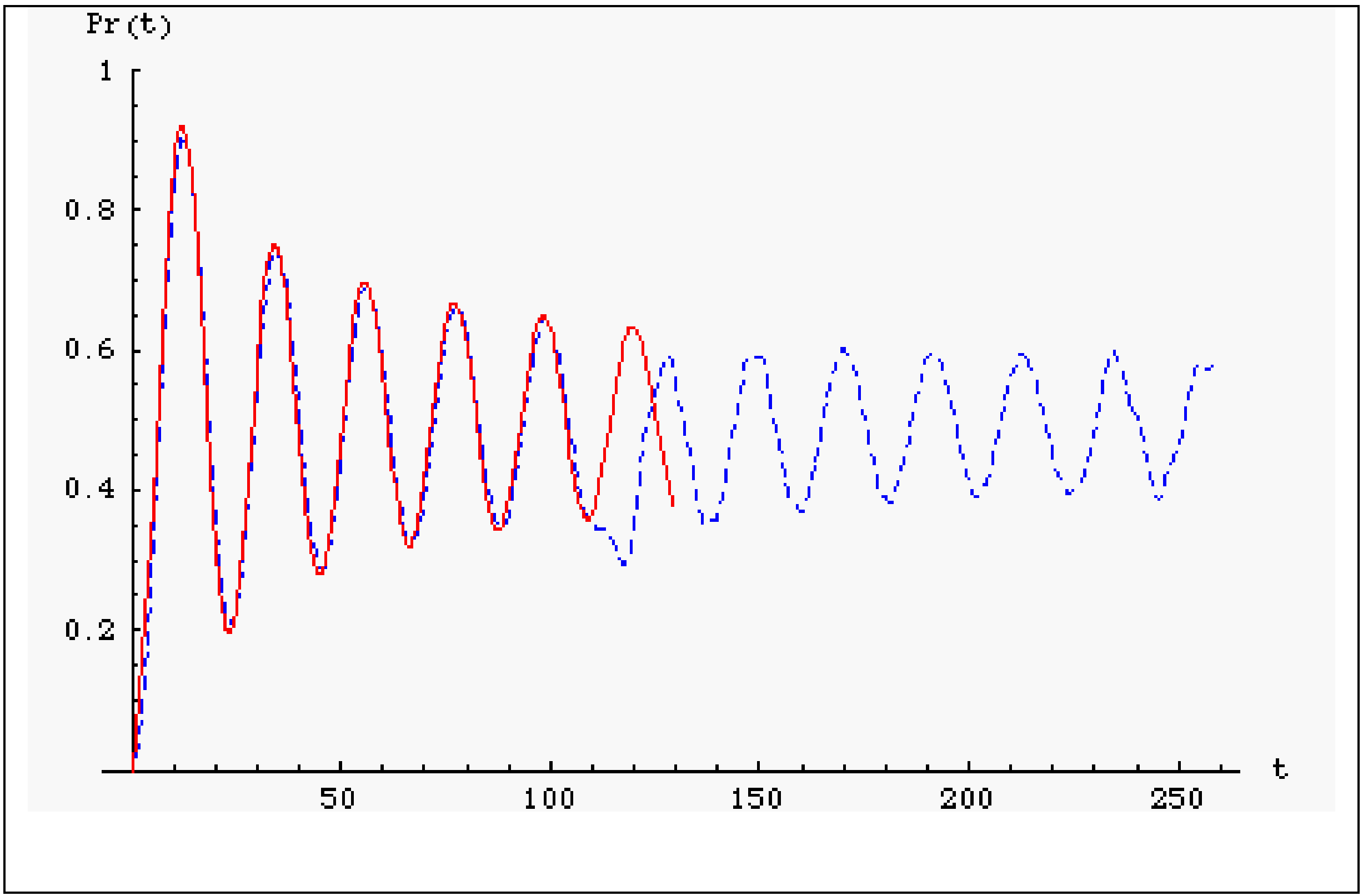

Figure 3

*The example shown here of* 2.39 (the solid line) *as an approximation to* 2.36 (the dashed line) *corresponds to the following choice of the parameters:*

$\mu = 6$;

$s = 2^{\mu+1} + 1 = 129$ *(with this choice, there are, in the superposition 2.19, terms corresponding to up to* $2^{\mu} = 64$ *queries of the oracle* $A$ *);*

$\lambda = 3 \cdot \pi/8$ *(with this choice [7], the average speed* $\langle \psi(t) | Q | \psi(t) \rangle / t$ *of the cursor is close to* 1 *for* $s >> 1$ *and for* $t < s$ *).*

The approximation 2.39 gives us control on the position $t_0$ of the first, and absolute, maximum of $\Pr(t)$:

$$t_0 = \frac{z_0}{2 \cdot \lambda \cdot \chi(\mu)} \approx \frac{z_0}{2 \cdot \lambda} 2^{\mu/2} \qquad (2.40)$$

where

$$z_0 \approx 3.518 \qquad (2.41)$$

is the position of the first zero, on the positive real axis, of the function $3 \cdot J_1(z) - J_3(z)$.

The height of the first, and absolute, maximum of $\Pr(t)$ is given by

$$\Pr(t_0) \approx \frac{1}{2} - \frac{1}{2}\left(J_0(z_0) - J_2(z_0)\right) \approx 0.92 \qquad (2.42)$$

## 3. Quantum subroutines.

The interest of the computational capabilities exhibited by the simple $XY$ model of figure 2 is strongly limited by the observation that, in order to observe the probability maximum 2.42, one needs a chain of cursor spins whose length $s$ grows exponentially with $\mu$. In this section we will show that this cost in terms of space can be made linear in $\mu$ by the use of quantum subroutines.

For every non negative integer $K$, we are goin to exhibit a quantum clocking mechanism able to apply $2^K$ times the transformation $\mathsf{BA}$ to the "register" qubits $\underline{\sigma}(1), \underline{\sigma}(2),..., \underline{\sigma}(\mu), \underline{\sigma}(\nu)$, by repeatedly using the same "piece of hardware" that applies $\mathsf{BA}$ just once.

This clocking mechanism will involve

$$s(K) = 4 \cdot K + 3 \tag{3.1}$$

"cursor" qubits $\underline{\tau}(1), \underline{\tau}(2),..., \underline{\tau}(s(K))$.

In order to keep track of the progress of the $2^K$ executions of the assigned "subroutine" $\mathsf{BA}$, there must be a subsystem (the "subroutine counter") having $2^K$ different states: it will be constructed in terms of $K$ qubits $\underline{\rho}(1), \underline{\rho}(2),..., \underline{\rho}(K)$.

For this additional set of spin 1/2 systems we will use, for j=1,..,K, notations such as $\underline{\rho}(j)=(\rho_1(j),\ \rho_2(j),\ \rho_3(j)) \equiv (\rho_x(j),\ \rho_y(j),\ \rho_z(j))$ and $\rho_\pm(j) = (\rho_x(j) \pm i \cdot \rho_y(j))/2$.

We will denote by $H_{counter}$ the $2^K$ dimensional state space of the "counter degrees of freedom".

The definition of the Hamiltonian operator on $H_{register} \otimes H_{cursor} \otimes H_{counter}$ will be given by an iterative scheme.

Set, for $i = 1,2,\ldots,s(K)-2$,

$$\begin{aligned} h_0(i,\ i+2) &= (A_{forward}(i) + B_{forward}(i+1)) \otimes 1_{counter} = \\ &= (\mathsf{A} \cdot \tau_+(i+1) \cdot \tau_-(i) + \mathsf{B} \cdot \tau_+(i+2) \cdot \tau_-(i+1)) \otimes 1_{counter} \end{aligned} \tag{3.2}$$

where $1_{counter}$ is the identity operator in $H_{counter}$. The operator $h_0(i; i+2)$ applies just once ($2^0 = 1$) the transformation $\mathsf{BA}$ to the register, while the cursor jumps from site $i$ to site $i+2$. This operator is represented graphically in figure 4.

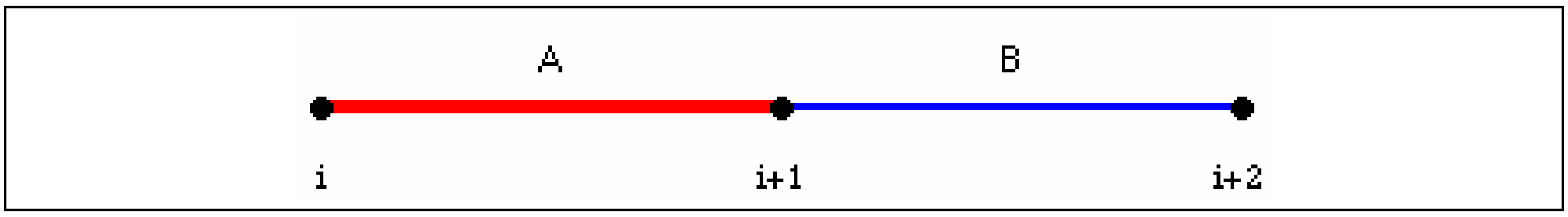

Figure 4
*"Do BA once"*

$$h_0(i,\ i+2) = \big(\mathsf{A}\cdot\tau_+(i+1)\cdot\tau_-(i) + \mathsf{B}\cdot\tau_+(i+2)\cdot\tau_-(i+1)\big)\cdot 1_{counter}$$

For $i = 1, 2, \ldots, s(K) - 6$ define

$$\begin{aligned} h_1(i, i+6) = {} & \rho_+(1)\cdot\tau_+(i+1)\cdot\tau_-(i) + \rho_x(1)\cdot\tau_+(i+2)\cdot\tau_-(i+1) + \\ & + h_0(i+2, i+4) + \\ & + \rho_-(1)\cdot\tau_+(i+6)\cdot\tau_-(i+4) + \rho_+(1)\cdot\tau_+(i+5)\cdot\tau_-(i+4) + \rho_-(1)\cdot\tau_+(i+1)\cdot\tau_-(i+5) \end{aligned} \tag{3.3}$$

A graphical representation of this term is given in figure 5.

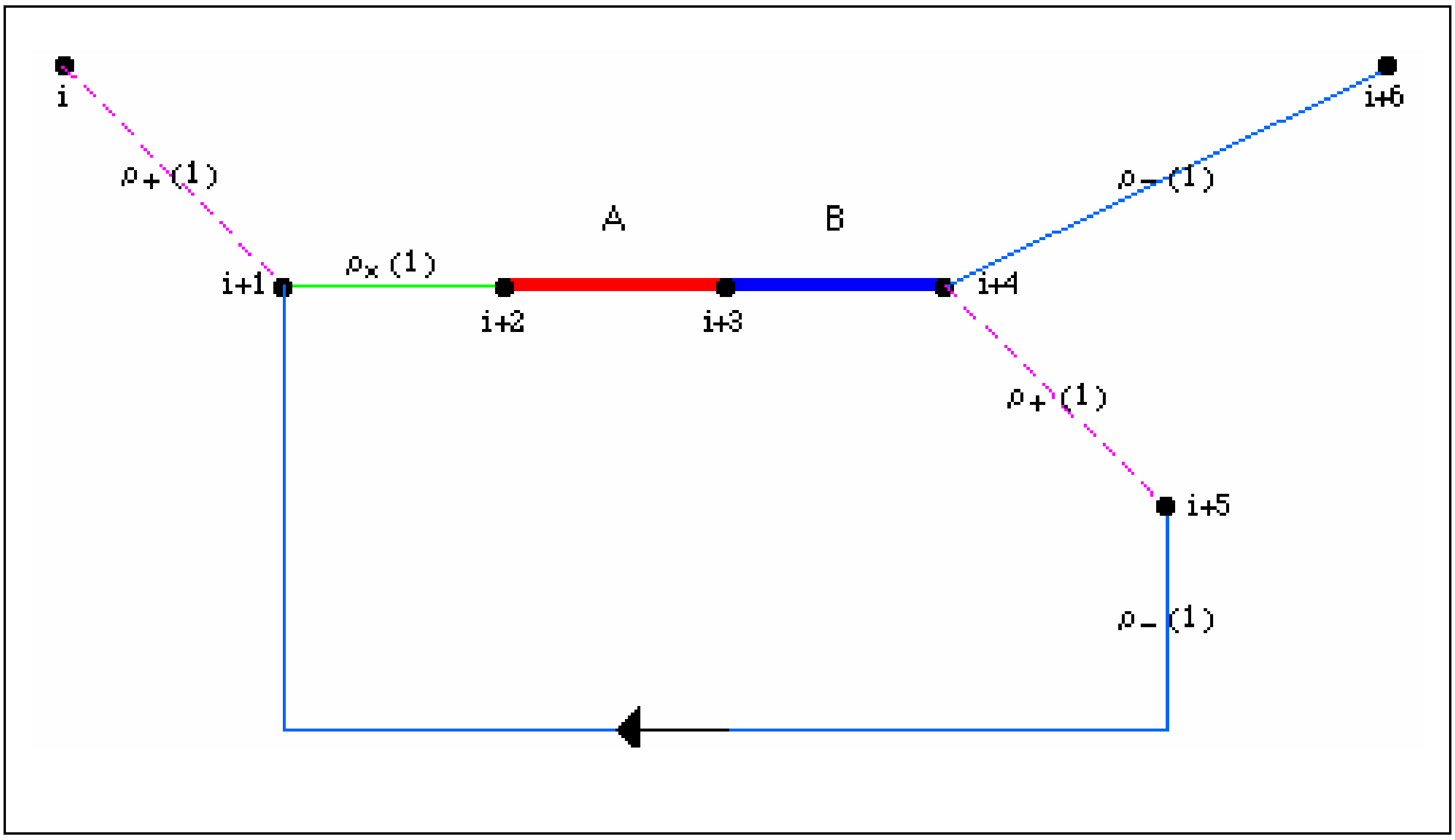

Figure 5
*"Do BA twice"*

The term $\rho_-(1)\cdot\tau_+(i+6)\cdot\tau_-(i+4) + \rho_+(1)\cdot\tau_+(i+5)\cdot\tau_-(i+4)$ in 3.3 is an example of the implementation of a conditional jump through the SWITCH primitive. The first addendum acts non-vanishingly only in the subspace belonging to the eigenvalue +1 of the controlling qubit $\rho_3(1)$ and sends the excitation of the cursor from $i+4$ to $i+6$; the second addendum, in turn, acts non-vanishingly only in the subspace belonging to the eigenvalue -1 of $\rho_3(1)$ and sends the excitation of the cursor from $i+4$ to $i+5$. Notice that in this implementation of the IF...THEN, ELSE... construct, the controlling bit $\rho_3(1)$ gets inverted.

The iteration step from $h_{j-1}$ to $h_j$ is given by:

$$
\begin{aligned}
&h_j(i, i+4\cdot j+2) = \rho_+(j)\cdot\tau_+(i+1)\cdot\tau_-(i) + \\
&+\rho_x(j)\cdot\tau_+(i+2)\cdot\tau_-(i+1) + \\
&+h_{j-1}(i+2, i+4\cdot j) + \\
&+\rho_-(j)\cdot\tau_+(i+4\cdot j+2)\cdot\tau_-(i+4\cdot j) + \\
&+\rho_+(j)\cdot\tau_+(i+4\cdot j+1)\cdot\tau_-(i+4\cdot j) + \\
&+\rho_-(j)\cdot\tau_+(i+1)\cdot\tau_-(i+4\cdot j+1)
\end{aligned}
\tag{3.4}
$$

and is represented in figure 6.

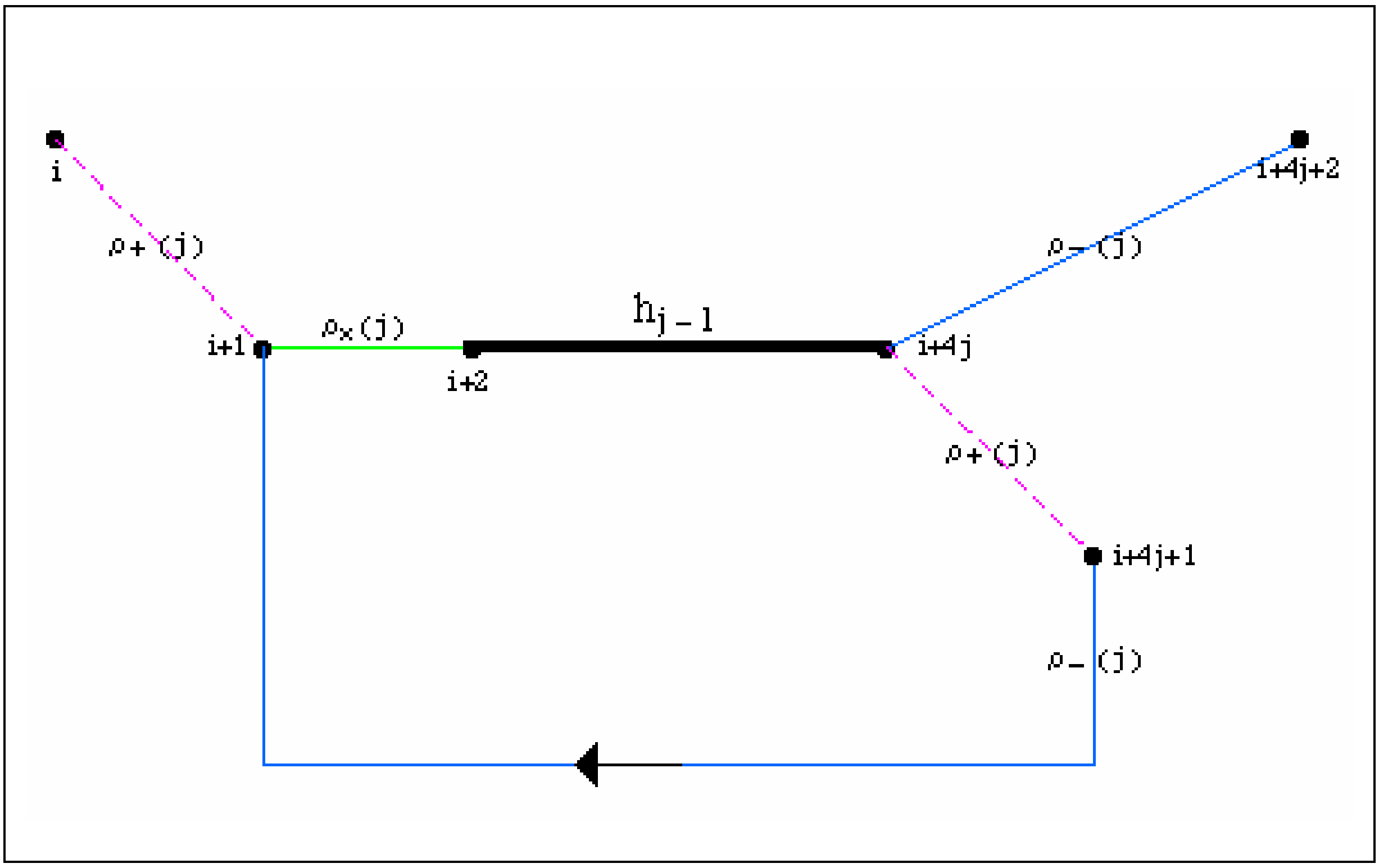

Figure 6

*"Do BA $2^j$ times while the cursor moves from $i$ to $i+s(j)-1$"*

For a fixed value of the positive integer $K$ we define the forward part of the Hamiltonian as

$$H_{forward}(K) = h_K(1, s(K)) = h_K(1, 4\cdot K+3) \tag{3.5}$$

and the Hamiltonian as

$$H(K) = H_{forward}(K) + H_{backward}(K) = H_{forward}(K) + H_{forward}(K)^* \tag{3.6}$$

We study, in this section, the Schrödinger equation

$$i \cdot \frac{d}{dt} \left| \psi(t) \right\rangle = -\frac{\lambda}{2} \cdot H(K) \left| \psi(t) \right\rangle \tag{3.7}$$

under the initial condition

$$\begin{aligned} \left| \psi(0) \right\rangle = & \left| \sigma_1(1) = 1, \ldots, \sigma_1(\mu) = 1, \sigma_1(\nu) = -1 \right\rangle \otimes \\ & \otimes \left| \tau_3(1) = 1, \tau_3(2) = -1 \ldots, \tau_3(s(K)) = -1 \right\rangle \otimes \\ & \otimes \left| \rho_3(1) = -1, \ldots, \rho_3(K) = -1 \right\rangle \equiv \left| \varphi_1 \right\rangle \end{aligned} \tag{3.8}$$

As in section 2, the problem 3.7, 3.8 is extremely easy to solve because of the conservation laws

$$\left[ H(K), \sigma_1(\nu) \right] = 0 \tag{3.9}$$

$$\left[ H(K), \sum_{j=1}^{s(K)} \tau_3(j) \right] = 0 \tag{3.10}$$

$$\left[ H(K), P(K) \right] = 0 \tag{3.11}$$

The operator $P(K)$ in 3.11 is the projector

$$P(K) = \sum_{j=1}^{p(K)} \left| \varphi_j \right\rangle \left\langle \varphi_j \right| \tag{3.12}$$

on the subspace spanned by the $p(K) = 2^{K+3} - 5$ orthonormal vectors defined by

$$\left| \varphi_1 \right\rangle = \left| \psi(0) \right\rangle \tag{3.13}$$

$$\left| \varphi_j \right\rangle = H_{forward}(K) \left| \varphi_{j-1} \right\rangle \quad \text{for} \quad j = 2, 3, \ldots, p(K) = 2^{K+3} - 5 \tag{3.14}$$

Because of the above considerations, the solution of 3.7, 3.8 will be of the form

$$\left| \psi(t) \right\rangle = \sum_{j=1}^{p(K)} c(t, j; p(K)) \left| \varphi_j \right\rangle \tag{3.15}$$

where, following the line of reasoning leading to 2.24, it is easy to show that

$$c(t, j; p(K)) = \frac{2}{p(K)+1} \sum_{n=1}^{s} \exp\left[ i \cdot \lambda \cdot t \cdot \cos(\vartheta(n; p(K))) \right] \cdot sin(\vartheta(n; p(K))) \cdot sin(j \cdot \vartheta(n; p(K))) \tag{3.16}$$

A full understanding of the solution 3.15 requires the analysis of the states $\left|\varphi_j\right\rangle$, for $j = 1,2,\ 3,\dots,p(K)$.

Because of 3.9, all of them are eigenstates of $\sigma_1(\nu)$ belonging to the eigenvalue -1. We will, from now on, omit the explicit reference to this fact, using the shorthand notation, for $\mathbf{x} \in \{-1,1\}^{\mu}$,

$$\left|\sigma_1(1) = x_1,\dots,\ \sigma_1(\mu) = x_\mu, \sigma_1(\nu) = -1\right\rangle \equiv \left|\sigma_1(1) = x_1,\dots,\sigma_1(\mu) = x_\mu\right\rangle \equiv |\mathbf{x}\rangle_1 \qquad (3.17)$$

Because of 3.10, each of the states $\left|\varphi_j\right\rangle$ is an eigenstate of the operator ("position of the cursor" or, with reference to the intuition developed in section 2, "position of the clocking quantum walk")

$$Q = \sum_{j=1}^{s(K)} j \cdot \frac{1 + \tau_3(j)}{2} \qquad (3.18)$$

This fact justifies a notation such as

$$\left|\tau_3(1) = -1, \tau_3(2) = -1,\dots, \tau_3(i) = +1,\dots,1,\ \tau_3(s(K)) = -1\right\rangle \equiv |Q = i\rangle \qquad (3.19)$$

Using these notations we will write, for instance

$$\left|\varphi_1\right\rangle = |\mathbf{1}\rangle_1 \otimes |Q = 1\rangle \otimes \left|\rho_3(1) = -1,\dots,\rho_3(K) = -1\right\rangle \equiv |\mathbf{1}\rangle_1 \otimes |Q = 1\rangle \otimes \left|\boldsymbol{\rho}_3 = \mathbf{-1}\right\rangle \qquad (3.20)$$

Each of the vectors $\left|\varphi_j\right\rangle$ will be, furthermore, a simultaneous eigenvector of each of the operators $\boldsymbol{\rho}_3 = (\rho_3(1),\rho_3(2),\dots,\rho_3(K))$. Calling $\mathbf{r}_j \in \{-1,1\}^K$ the collection of the eigenvalues to which $\left|\varphi_j\right\rangle$ belongs, we will write

$$\left|\varphi_j\right\rangle = \mathsf{A}^{\varepsilon_j}(\mathsf{BA})^{n_j}|\mathbf{1}\rangle_1 \otimes \left|Q = q_j\right\rangle \otimes \left|\boldsymbol{\rho}_3 = \mathbf{r}_j\right\rangle \qquad (3.21)$$

The explicit iterative algorithm by which $\varepsilon_j, n_j, q_j, \mathbf{r}_j$ can be computed is strictly parallel to the iteration procedure of figures 4 and 6.

For the discussion that follows it is sufficient to give here the explicit expressions of the exponents $\varepsilon_j$ and $n_j$.

$$\varepsilon_j = \begin{cases} 1 \ if\ j \in \{j_1, j_2,\dots,j_{2^K}\} \\ \\ 0 \qquad\qquad \text{otherwise} \end{cases} \qquad (3.22)$$

where, for $i = 1,2,\dots,2^K$,

$$j_i = 2 \cdot K + 2 + 5 \cdot (i-1) + 3 \cdot \sum_{x=1}^{i-1} \mathrm{e}_2(x) = 2 \cdot K + 2 + 5 \cdot (i-1) + 3 \cdot \sum_{h=1}^{K-1} \left\lfloor (i-1) / 2^{K-h} \right\rfloor \qquad (3.23)$$

In (3.23) we have indicated by $\mathrm{e}_2(x)$ the exponent of the prime factor 2 in the factorization of the positive integer $x$, and by $\lfloor y \rfloor$ the integer part of the positive real number $y$.

Let's focus our attention on the states $\left|\varphi_{j_1}\right\rangle, \left|\varphi_{j_2}\right\rangle, \ldots, \left|\varphi_{j_{2^K}}\right\rangle$.

$$\left|\varphi_{j_1}\right\rangle = \mathsf{A}|\mathbf{1}\rangle_1 \otimes |Q = 2 \cdot K + 2\rangle \otimes |\rho_3(1) = -1, \ldots, \rho_3(K) = -1\rangle \equiv |\mathbf{1}\rangle_1 \otimes |Q = 2 \cdot K + 2\rangle \otimes \left|N_{\rho_3} = 1\right\rangle, \qquad (3.24)$$

where $j_1 = 2 \cdot K + 2$.

In (3.24) we have given a numerical meaning to the content of the subroutine counter by defining the operator

$$N_{\rho_3} = 1 + \sum_{y=1}^{K} \frac{1 + \rho_3(y)}{2} 2^{y-1} \qquad (3.25)$$

In all the predecessors $|\varphi_1\rangle, |\varphi_2\rangle, \ldots, \left|\varphi_{j_1 - 1}\right\rangle$ of $\left|\varphi_{j_1}\right\rangle$ the register is in its initial state $|\mathbf{1}\rangle_1$. The immediate successor of $\left|\varphi_{j_1}\right\rangle$ is

$$\left|\varphi_{j_1+1}\right\rangle = \mathsf{BA}|\mathbf{1}\rangle_1 \otimes |Q = 2 \cdot K + 3\rangle \otimes \left|N_{\rho_3} = 1\right\rangle \qquad (3.26)$$

In all of the states $\left|\varphi_{j_1+1}\right\rangle, \ldots, \left|\varphi_{j_2-1}\right\rangle$ the register remains in the state $\mathsf{BA}|\mathbf{1}\rangle_1$; the content of the register changes only at step $j_2$, where it is

$$\left|\varphi_{j_2}\right\rangle = \mathsf{ABA}|\mathbf{1}\rangle_1 \otimes |Q = 2 \cdot K + 7\rangle \otimes \left|N_{\rho_3} = 2\right\rangle \qquad (3.27)$$

At each of the steps $j_i$ the state of the register gets acted upon by an additional $\mathsf{A}$ and at step $j_i + 1$ by an additional $\mathsf{B}$. In steps from $j_i + 2$ to $j_{i+1} - 1$ the state of the register remains unaltered.

The content of the register becomes $(\mathsf{BA})^{2^K} |\mathbf{1}\rangle_1$ for the first time at step $j_{2^K} + 1 = p(K) - K$ and such remains until the last step $p(K)$.

The exponent $n_j$ in 3.21 is therefore equal to the number of "non-trivial" steps $j_i$ that precede step $j$:

$$n_j = \left|\left\{1 \le i \le 2^K : j_i < j\right\}\right| \qquad (3.28)$$

It is, therefore

$$
\begin{aligned}
&n_j = 0 \text{ for } j < 2\cdot K + 2\\
&n_{2\cdot K+3} = 1\\
&n_j = 2^K \text{ for } j \geq p(K) - K
\end{aligned}
\tag{3.29}
$$

For $2\cdot K + 3 \leq j \leq p(K) - K$, $n_j$ grows in an approximately linear way because of the inequalities

$$2\cdot K + 2 + 5\cdot(i-1) + 3\cdot\left(1 - \frac{1}{2^{L(i)}}\right)\cdot(i-1) - 3\cdot L(i) \leq j_i \qquad (3.30_a)$$

$$j_i \leq 2\cdot K + 2 + 5\cdot(i-1) + 3\cdot\left(1 - \frac{1}{2^{L(i)}}\right)\cdot(i-1) \qquad (3.30_b)$$

with $L(i) = \lfloor \log_2(i-1) \rfloor$, which easily follow from 3.23 and from the fact that $x - 1 < \lfloor x \rfloor \leq x$.

This justifies the approximation

$$
n_j \approx \begin{cases}
0 \text{ for } 1 \leq j \leq 2\cdot K + 2\\
1 + \dfrac{2^K - 1}{p(K) - 3\cdot K - 3}\left(j - (2\cdot K + 3)\right) \text{ for } 2\cdot K + 3 \leq j \leq p(K) - K\\
2^K \text{ for } p(K) - K \leq j \leq p(K)
\end{cases}
\tag{3.31}
$$

that we shall use in what follows.

Following the same line of reasoning that lead from 2.19 to 2.36, we can conclude that, in the state 3.15, the probability $\Pr(t)$ of finding the register in the state $|\mathbf{a}\rangle_3$ is given by:

$$\Pr(t) = \sum_{j=1}^{p(K)} \alpha_{n_j}(\mu)^2 \cdot \left|c(t, j; p(K))\right|^2 \approx \qquad (3.32_a)$$

$$\approx \sum_{j=1}^{p(K)} \left(\sin\left(\chi(\mu)\cdot\left(2\cdot n_j + 1\right)\right)\right)^2 \cdot \frac{4\cdot j^2}{(\lambda\cdot t)^2} J_j(\lambda\cdot t)^2 \approx \qquad (3.32_b)$$

$$\approx \sum_{j=1}^{p(K)} \left(\sin\left(\chi(\mu)\cdot\frac{j}{4}\right)\right)^2 \cdot \frac{4\cdot j^2}{(\lambda\cdot t)^2} J_j(\lambda\cdot t) \approx$$

$$\approx \int_0^{\lambda\cdot t} \frac{4\cdot x^2}{\pi\cdot(\lambda\cdot t)^2\cdot\sqrt{(\lambda\cdot t)^2 - x^2}} \left(\sin\left(\chi(\mu)\cdot\frac{x}{4}\right)\right)^2 dx =$$

$$= \frac{1}{2} - \frac{1}{2}\left(J_0\left(\chi(\mu)\cdot\lambda\cdot t/2\right) - J_2\left(\chi(\mu)\cdot\lambda\cdot t/2\right)\right) \qquad (3.32_c)$$

Figure 7 shows a plot of the exact expression $(3.32_a)$ of $\Pr(t)$ and of its approximation $(3.32_c)$. It corresponds, for the sake of comparison with figure 3 to the following choice of parameters: $\mu = 6$, $K = 6$ (with this choice, there are, in the superposition 3.15, terms corresponding to up to $2^\mu$ queries of the oracle $\mathsf{A}$ ), $\lambda = 3 \cdot \pi/8$ (with this choice [7], the computation proceeds at an average rate of one transition per unit of time in the sense that $\sum_{j=1}^{p(K)} j \cdot |c(t,j;p(K))|^2 \approx t$ for $0 < t < p(K)$).

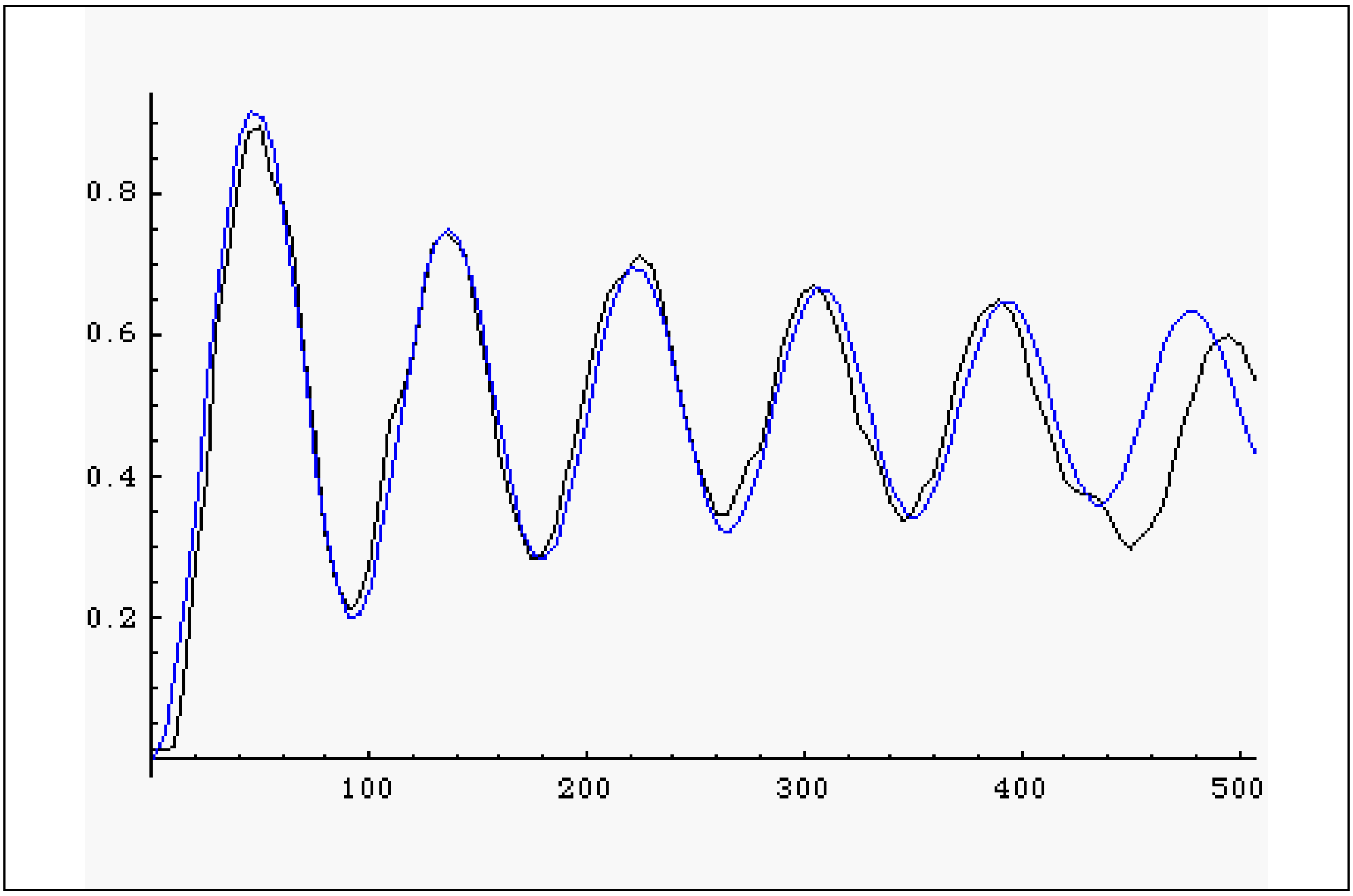

Figure 7
Both sides of the approximate equality
$$\Pr(t) \approx \frac{1}{2} - \frac{1}{2}\left(J_0(\chi(\mu) \cdot \lambda \cdot t/2) - J_2(\chi(\mu) \cdot \lambda \cdot t/2)\right) \text{ are plotted.}$$
As in figure 3, the approximate expression misses the abrupt phase change present in the exact result, which is due to reflection at the end of the computational path.

## 4. Equivalent "local" Hamiltonians.

The explicit expression, in terms of the register spins, of the "oracle" operator $\mathsf{A}$ defined in 2.2 is:

$$\mathsf{A} = I + (\sigma_1(\nu) - I) \cdot \prod_{i=1}^{\mu} \frac{I + a_i \cdot \sigma_3(i)}{2} \quad . \tag{4.1}$$

The analogous expression for the "estimator" operator $\mathsf{B}$ defined in 2.3 is

$$\mathsf{B} = I + \left(\sigma_1(\nu) - I\right) \cdot \prod_{i=1}^{\mu} \frac{I + \sigma_1(i)}{2}, \tag{4.2}$$

where $I$ is the identity operator in $H_{register}$.

In the Hamiltonian $H$ defined by 2.11 and in the Hamiltonian $H(K)$ defined by 3.6 there are, therefore, "non-local" terms such as $\mathsf{A} \otimes \tau_+(j+1) \cdot \tau_-(j)$ and $\mathsf{B} \cdot \tau_+(j+1) \cdot \tau_-(j)$ involving many-body interactions among two cursor spins and <u>all</u> the register spins.

This section is a brief digression on the analysis of the computational cost, in terms of space (additional qubits), (average) time, and probability (of ever finding the computation completed) involved in substituting such non-local terms with equivalent terms in which only interactions between two cursor spins and at most <u>one</u> register spin appear.

For the sake of definiteness we concentrate our attention, to start with, on the CNOT primitive:

$$C^{\mu}NOT(j, j+1) = \mathsf{CNOT} \otimes \tau_+(j+1) \cdot \tau_-(j) + hermitian\,conjugate, \tag{4.3}$$

where

$$\mathsf{CNOT} = I + \left(\sigma_1(\nu) - I\right) \cdot \prod_{i=1}^{\mu} \frac{I + \sigma_3(i)}{2}. \tag{4.4}$$

("Flip the $z$ component of the $\nu - th$ qubit iff all the $\mu$ input qubits point in the $+z$ direction, starting with the cursor in position $j$")

The case $\mu = 1$ of one controlling qubit has been studied in [4]. It involves the introduction of $s_1 = 6$ cursor qubits $\underline{\tau}$(j), $\underline{\tau}$(j+1),..., $\underline{\tau}$(j+5) and, supposing that the controlling qubit is $\sigma_3(1)$ and the controlled one is $\sigma_3(\nu)$, of the "local" Hamiltonian

$$
\begin{aligned}
c^1 not(j, j+5) = {} & \sigma_-(1)\cdot \tau_-(j)\cdot \tau_+(j+1)+ \\
& + \sigma_1(\nu)\cdot \tau_-(j+1)\cdot \tau_+(j+2)+ \\
& + \sigma_+(1)\cdot \tau_-(j+2)\cdot \tau_+(j+5)+ \\
& + \sigma_+(1)\cdot \tau_-(j)\cdot \tau_+(j+3)+ \\
& + I\cdot \tau_-(j+3)\cdot \tau_+(j+4)+ \\
& + \sigma_-(1)\cdot \tau_-(j+4)\cdot \tau_+(j+5)+ \\
& + hermitian\ \ conjugate\ .
\end{aligned}
\tag{4.5}
$$

A graphical representation of 4.5 is given in figure 8.

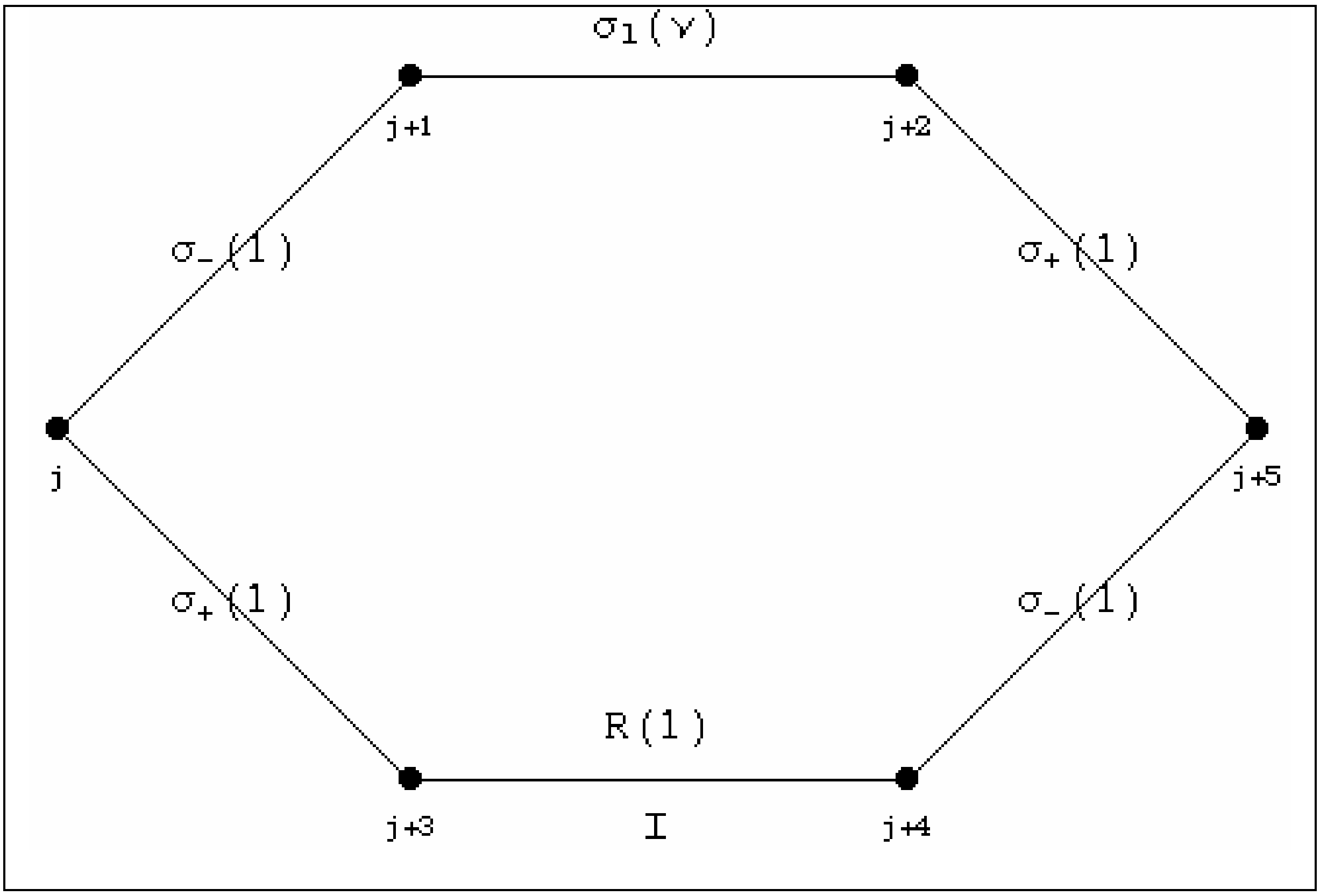

Figure 8
$c^1 not(j, j+5)$
This is a streamlined version of figure 8 of Reference [4]

The term $R(j+3, j+4) = I\cdot \tau_-(j+3)\cdot \tau_+(j+4)$ in 4.5, represented as $R(1)$ in figure 8, plays the role of a delay line of length 1. It makes the length $T_1 = 4$ of the computation independent of the input word in the sense that an initial state of the form $|\sigma_3(1) = 1\rangle \otimes |\sigma_3(\nu) = z_\nu\rangle |Q = j\rangle$ has the same <u>number</u> of logical successors

$$
\begin{aligned}
&|\sigma_3(1) = -1\rangle \otimes |\sigma_3(\nu) = \ \ z_\nu\rangle |Q = j+1\rangle \\
&|\sigma_3(1) = -1\rangle \otimes |\sigma_3(\nu) = -z_\nu\rangle |Q = j+2\rangle \\
&|\sigma_3(1) = +1\rangle \otimes |\sigma_3(\nu) = -z_\nu\rangle |Q = j+5\rangle
\end{aligned}
\tag{4.6}
$$

as an initial state of the form $|\sigma_3(1) = -1\rangle \otimes |\sigma_3(\nu) = z_\nu\rangle |Q = j\rangle$, which has the successors

$$
\begin{aligned}
&|\sigma_3(1) = +1\rangle \otimes |\sigma_3(\nu) = z_\nu\rangle |Q = j+3\rangle \\
&|\sigma_3(1) = +1\rangle \otimes |\sigma_3(\nu) = z_\nu\rangle |Q = j+4\rangle \\
&|\sigma_3(1) = -1\rangle \otimes |\sigma_3(\nu) = z_\nu\rangle |Q = j+5\rangle
\end{aligned}
\tag{4.7}
$$

Figure 9 shows the iteration step leading from $c^{\mu-1}not$ to $c^{\mu}not$ through the introduction of the additional controlling qubit $\sigma_3(\mu)$. The length of each computation increases from the previous value $T_{\mu-1}$ to

$$T_\mu = T_{\mu-1} + 2 = 2 \cdot (\mu + 1) \tag{4.8}$$

The number of cursor qubits increases, because also of the delay line $R(T_{\mu-1})$, from the previous value $s_{\mu-1}$ to

$$s_\mu = 2 + s_{\mu-1} + T_{\mu-1} = (\mu+1) \cdot (\mu+2) \tag{4.9}$$

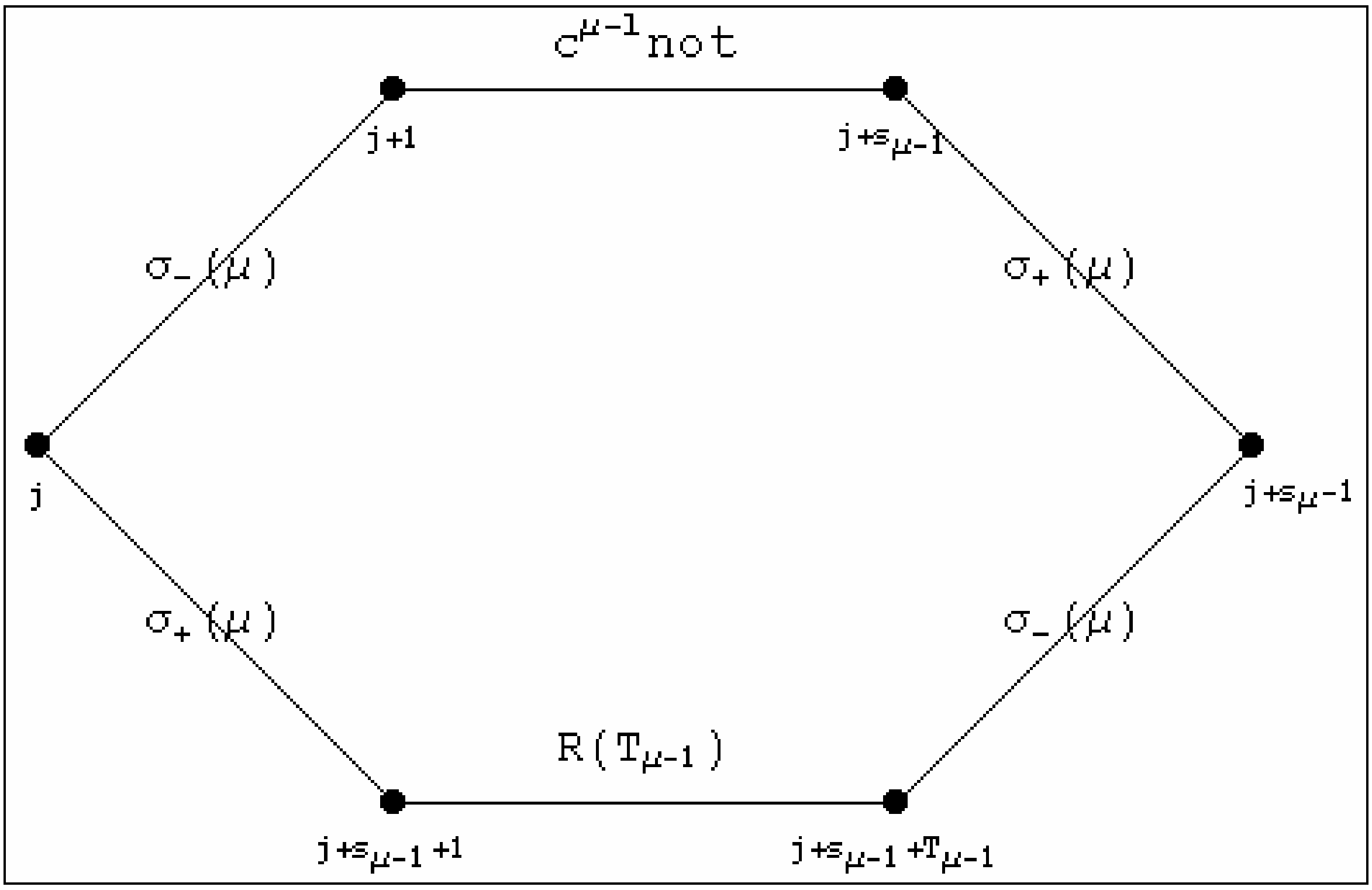

Figure 9

$$c^{\mu-1}not \to c^{\mu}not$$

The iteration step $c^{\mu-1}not \to c^{\mu}not$ is explicitly given by

$$\begin{aligned} c^{\mu}not\left(j, j+s_{\mu}-1\right) = {} & \sigma_{-}(\mu)\cdot \tau_{-}(j)\cdot \tau_{+}(j+1) + \\ & + c^{\mu-1}not\left(j+1, j+s_{\mu-1}\right) + \\ & + \sigma_{+}(\mu)\cdot \tau_{-}\left(j+s_{\mu-1}\right)\cdot \tau_{+}\left(j+s_{\mu}-1\right) + \\ & + \sigma_{+}(\mu)\cdot \tau_{-}(j)\cdot \tau_{+}\left(j+s_{\mu-1}+1\right) + \\ & + \sum_{k=1}^{T_{\mu-1}-1} \tau_{-}\left(j+s_{\mu-1}+k\right)\cdot \tau_{+}\left(j+s_{\mu-1}+k+1\right) + \\ & + \sigma_{-}(\mu)\cdot \tau_{-}\left(j+s_{\mu-1}+T_{\mu-1}\right)\cdot \tau_{+}\left(j+s_{\mu}-1\right) + \\ & + \textit{hermitian conjugate} \end{aligned} \tag{4.10}$$

Of particular conceptual relevance is the case $\mu = 2$ of the CCNOT (or TOFFOLI) primitive (figure 10), known to be a universal reversible primitive.

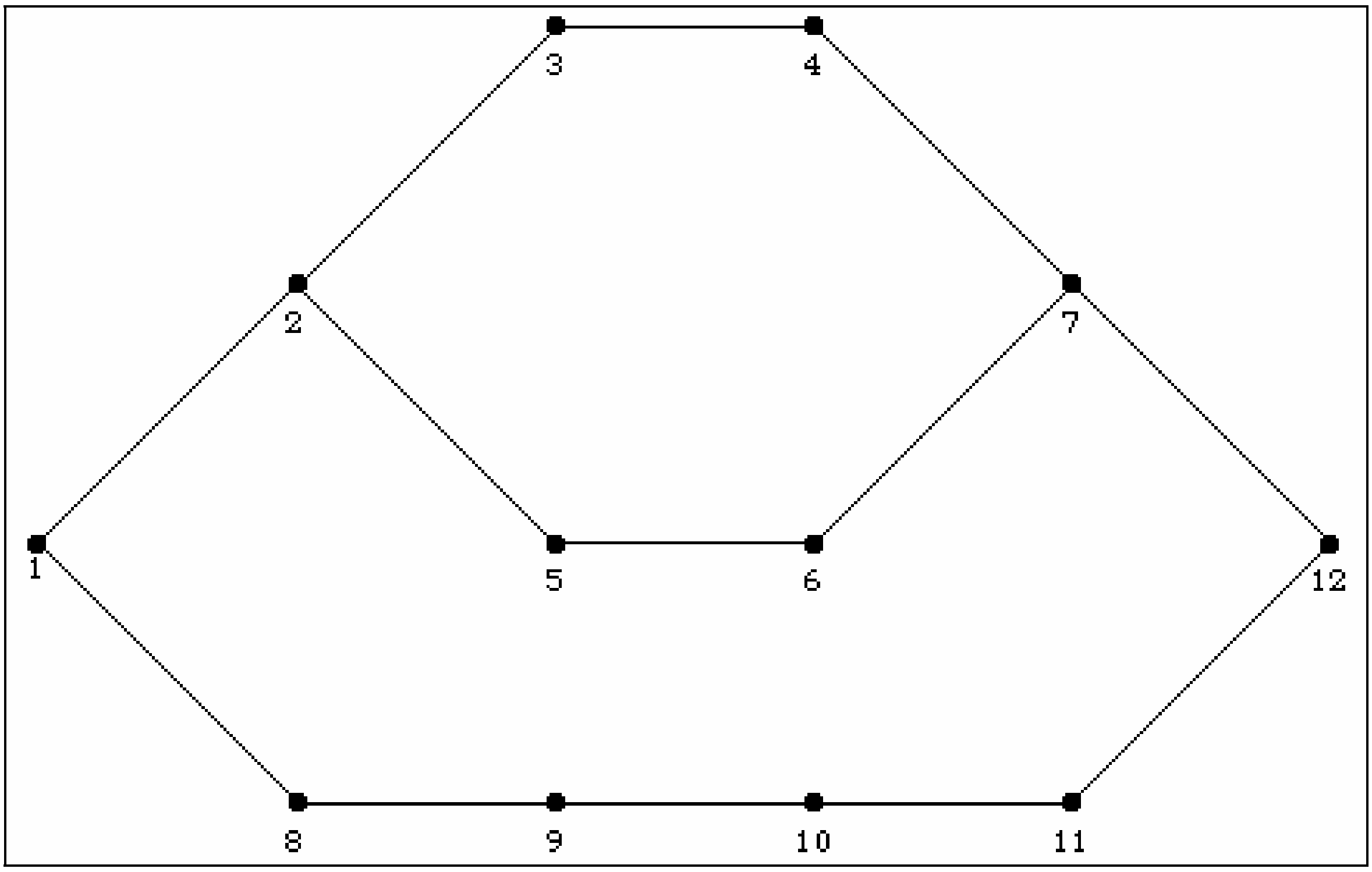

Figure 10

$c^{2}not(1,12)$

*Links $(1,2)$ and $(2,3)$ correspond to terms in the forward part of the Hamiltonian which act non-trivially (by flipping them) only on states having the two controlling bits "up". The term of link $(3,4)$ acts on such states by flipping the controlled bit. If one of the controlling bits is*

*"down", only the delay lines are* $(5,6)$ *and* $(8,11)$ *are active, instead. The other links correspond to terms that restore the controlling bits to their initial values.*

Figure 11 corresponds to the solution of the Schrödinger equation

$$i \cdot \frac{d}{dt}|\psi(t)\rangle = -\frac{\lambda}{2} c^2not(1,\ 12)|\psi(t)\rangle \tag{4.11}$$

under an initial condition of the form $|\psi(0)\rangle = |\sigma_3(1) = z_1, \sigma_3(2) = z_2\rangle \otimes |\sigma_3(\nu) = z_\nu\rangle \otimes |Q = 1\rangle$.

It gives, as a function of time, the probability $|c(t, T_2; T_2)|^2$ of finding the cursor in the state $|Q = s_2\rangle$ and the register in the state $|\sigma_3(1) = z_1, \sigma_3(2) = z_2\rangle \otimes |\sigma_3(\nu) = (1 - 2 \cdot \delta_{1,z_1} \cdot \delta_{1,z_2}) \cdot z_\nu\rangle$.

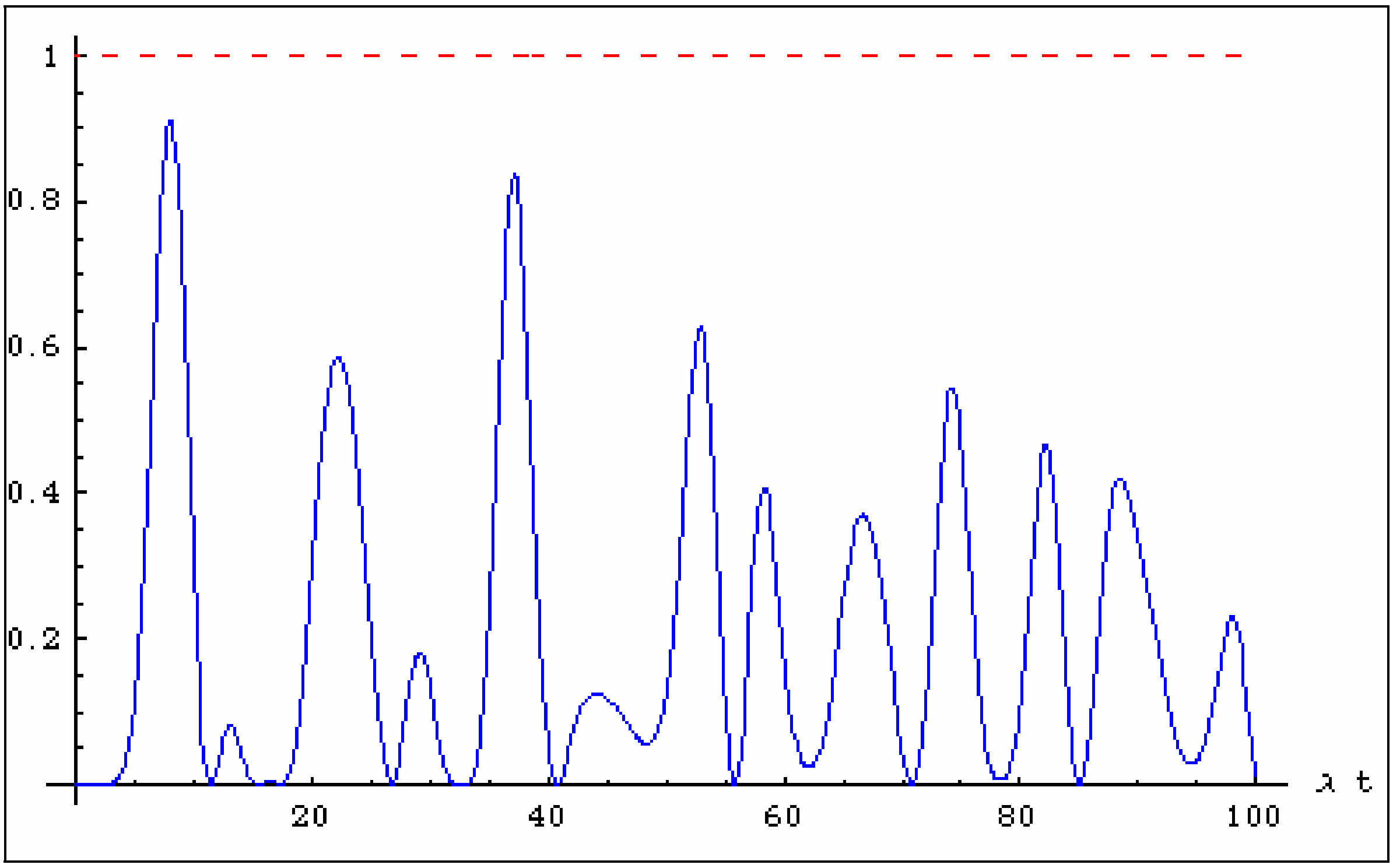

Figure 11
*Probability of finding the computation of CCNOT completed.*
*Notice that at every time this probability is strictly smaller than* 1.

**5. Numerical examples.**

The analysis of CNOT in terms of three-body interactions presented in section 4 can be extended, of course, to the "estimator" B: one has just to substitute $\sigma_3(i)$ with $\sigma_1(i)$ in 4.4. and, correspondingly (in figure 9 and in 4.10) the raising and lowering operators for the $z$ components of the register spins $\sigma_{\pm}(j) = \left(\sigma_x(j) \pm i \cdot \sigma_y(j)\right)/2$ with the raising and lowering operators $\left(\sigma_z(j) \mp i \cdot \sigma_y(j)\right)/2$ for the $x$ components.

The cost of writing B in terms of at most three-body interactions is, therefore, quadratic in terms of space, because $s_\mu = (\mu + 1) \cdot (\mu + 2)$.

This cost is linear in terms of time, as measured, say, by the position of the first maximum of $\left|c\left(t, T_\mu; T_\mu\right)\right|^2$, which grows as an approximately linear function of $T_\mu = 2 \cdot (\mu + 1)$.

There is, in addition, as shown in [7] and exemplified in figure 11 a non-trivial cost in terms of probability because of the upper bound

$$\left|c\left(t, T_\mu; T_\mu\right)\right|^2 \leq const / \left(T_\mu\right)^{2/3}. \tag{5.1}$$

We refer the reader to eq. 1.3 of Reference 7 for an argument showing that the above probability cost <u>can</u> be compensated for by the addition of a chain of telomeric sites of length proportional to $T_\mu = 2 \cdot (\mu + 1)$; in the examples that follow we neglect, for simplicity, this additional linear space cost.

In this section we will suppose, essentially for notational convenience, that the "oracle" A has been implemented, in a way analogous to the "estimator" B, in terms of three-body interactions through the obvious substitution of $\sigma_3(i)$ with $a_i \cdot \sigma_3(i)$ in 4.4 and, correspondingly, of $\sigma_{\pm}(j)$ with $\left(\sigma_x(j) \pm i \cdot a_j \cdot \sigma_y(j)\right)/2$ in figure 9 and in 4.10.

Only minor changes would be needed if A had a different time cost in terms of the number $T_{\mathsf{A}}$ of logical successors of an initial condition: the only point to stress is that $T_{\mathsf{A}}$ would appear in the analysis of the probabilistic aspects of the algorithm, for instance in determining the length of a telomeric chain required to achieve a given level of probability.

Figure 12 shows the simplest non-trivial example (corresponding to $\mu = 3$ and $K = 1$) of the kind of systems that emerge from the above considerations.

The dots in figure 12 refer to the cursor spins τ . The links in the left and right "handles", such as $(1,2)$ and $(41,42)$, represent interactions between the corresponding cursor spins mediated by the subroutine counter spin ρ .

The ascending or descending links in A describe interactions between cursor spins mediated by the various $\left(\sigma_x(j) \pm i \cdot a_j \cdot \sigma_y(j)\right)/2$, $j = 1, \ldots, \mu$ .

In B there appear, instead, with the same role, $\left(\sigma_z(j) \mp i \cdot \sigma_y(j)\right)/2$ .

The output qubit $\sigma_1(\nu)$ appears in the "top" links $(6,7)$ and $(25,26)$.

All the other interactions, in the delay lines, are pure $XY$ interactions between cursor spins.

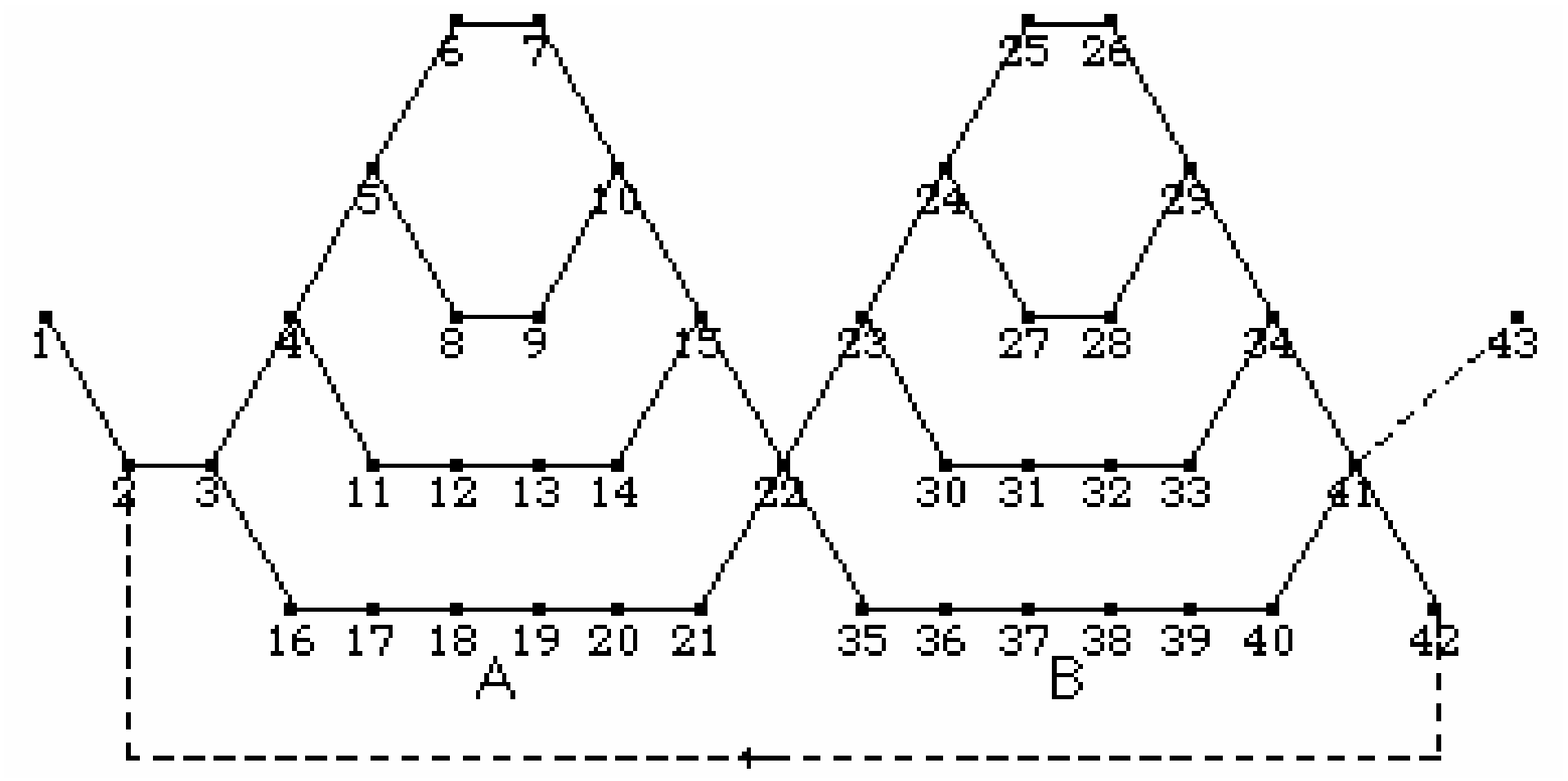

Figure 12

*"Do BA twice" on a register of* $\mu = 3$ *qubits as described in terms of three-body interactions*

We wish to describe in some detail the time evolution $|\psi(t)\rangle$ of such a system, starting from the initial condition $|\psi(0)\rangle = |\sigma_1(1) = 1, \sigma_1(2) = 1, \sigma_1(3) = 1, \sigma_1(4) = -1\rangle \otimes |Q = 1\rangle \otimes |\rho_3(1) = -1\rangle$.

Figure 13 describes the behaviour of the subroutine counter in terms of the expectation value $\langle \psi(t) | \rho_3(1) | \psi(t) \rangle$.

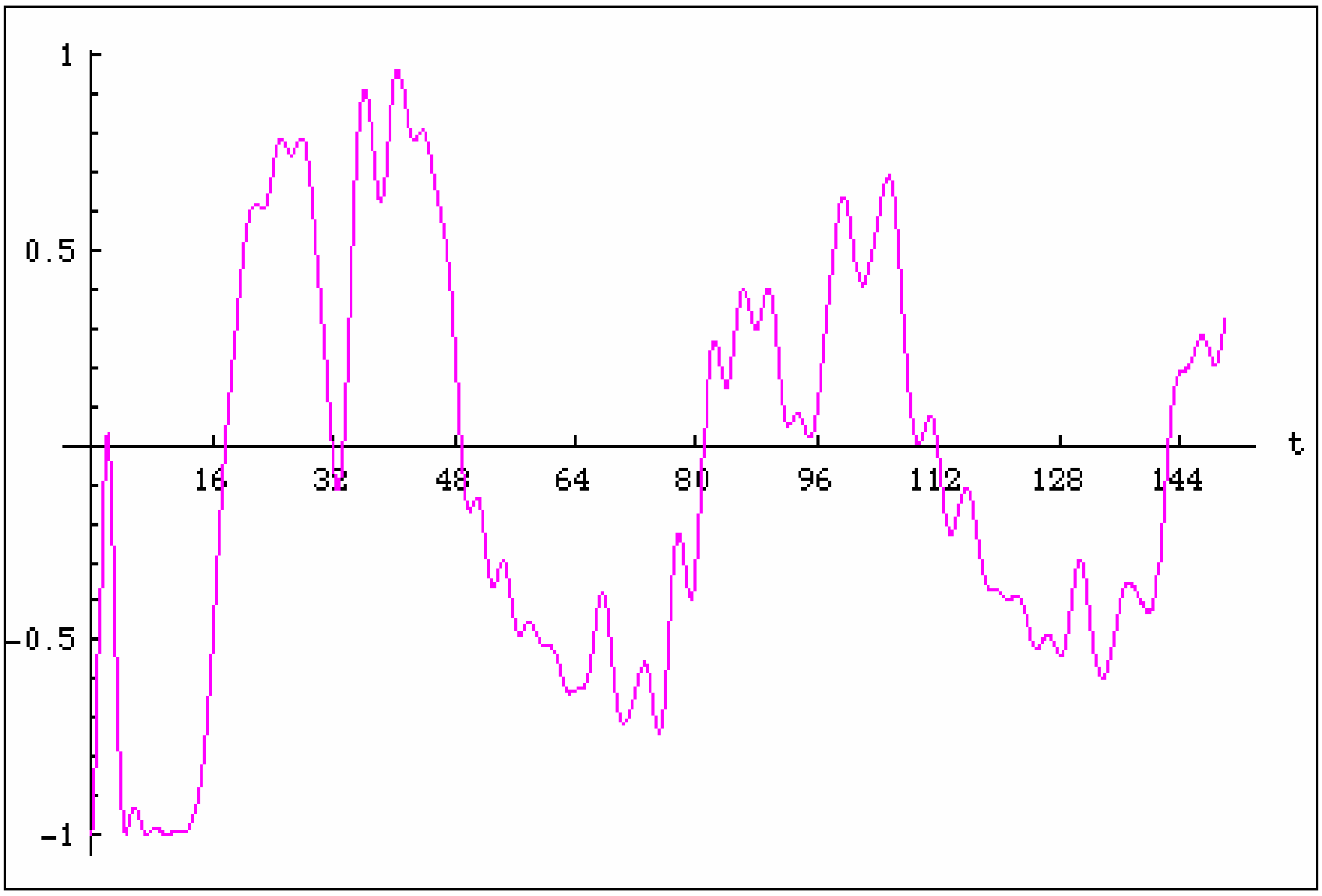

Figure 13

*$\langle \psi(t) | \rho_3(1) | \psi(t) \rangle$ as a function of $t$. We are using, as usual, $\lambda = 3 \cdot \pi/8$, so that the time coordinate can be read as the mean value of the number of computational steps performed.*

Figure 14 describes the behaviour of the cursor in terms of the expectation value $\langle \psi(t) | Q | \psi(t) \rangle$.

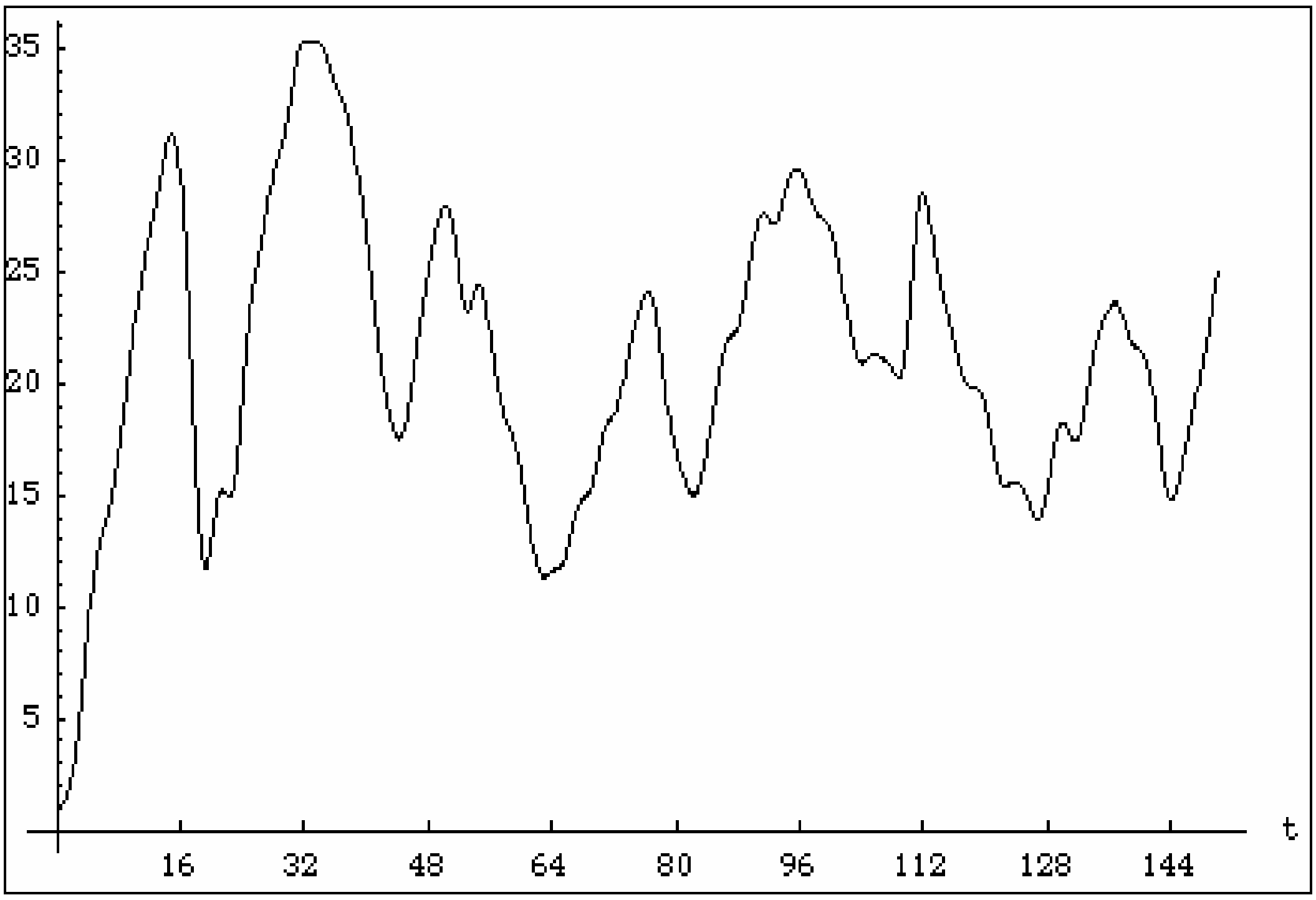

Figure 14

$\langle \psi(t)|Q|\psi(t)\rangle$ *as a function of time. Here and in figure* 13 *the ticks on the time axis call attention on the significant instants in which* AB *has been executed once and twice and on the instants in which the cursor gets reflected at the boundaries.*

Figure 15 shows the behaviour of the register in terms of the expectation value of the projector on the state $|\sigma_3(1)=a_1,\sigma_3(2)=a_2,\sigma_3(3)=a_3,\sigma_1(4)=-1\rangle\otimes|Q=43\rangle\otimes|\rho_3(1)=-1\rangle$.

In equivalent terms, what is represented, as a function of time, in figure 15 is the probability that the two following conditions are both satisfied: the clocking degrees of freedom take the values $Q=43$ and $\rho_3(1)=-1$ corresponding to the computation having been completed and the register is in the target state.

Figure 16 shows a less conventional usage of the same machine: at any time of your choice read the content of register only. In more precise terms, figure 16 gives a plot of the expectation value $\langle \psi(t)|P_{\mathbf{a}}|\psi(t)\rangle$ of the projector $P_{\mathbf{a}} = |\mathbf{a}\rangle_{33}\langle\mathbf{a}|\otimes 1_{cursor}\otimes 1_{counter}$ .

Notice that, in the case of Grover's algorithm, there is a definite advantage, offered by the quantum clocking mechanism studied here, in having a superposition of states in which BA has acted different numbers of times on the register: it is precisely the entanglement between the states of the register and the states of the quantum clock described by 3.15 that leads to the behaviour of the upper graph in figure 16.

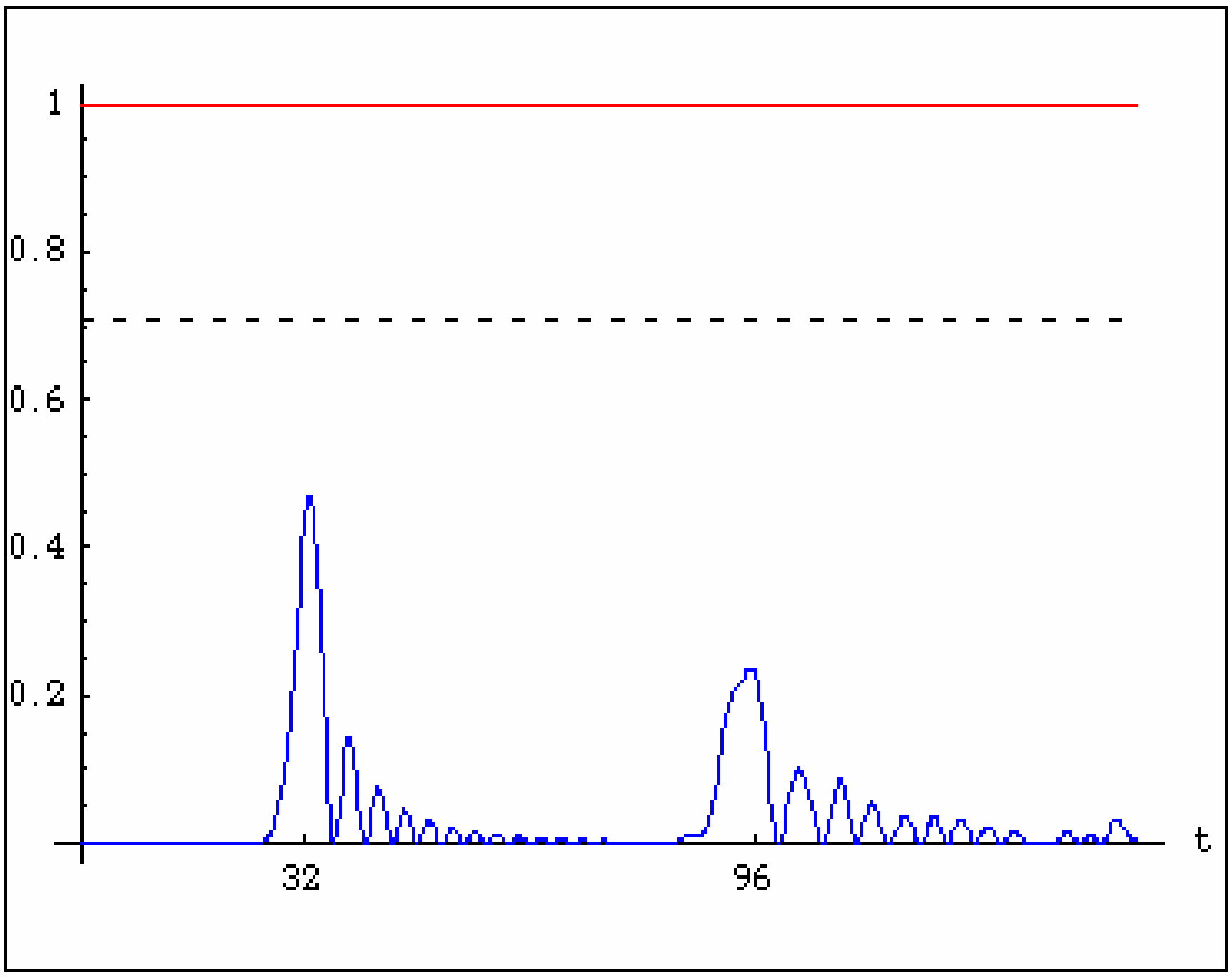

Figure 15

*Probability that* $(Q=43)\wedge(\rho_3(1)=-1)\wedge(\sigma_3(1)=a_1)\wedge(\sigma_3(2)\wedge a_2)\wedge(\sigma_3(3)=a_3)$

*as a function of time. The dashed line is the upper bound 5.1*

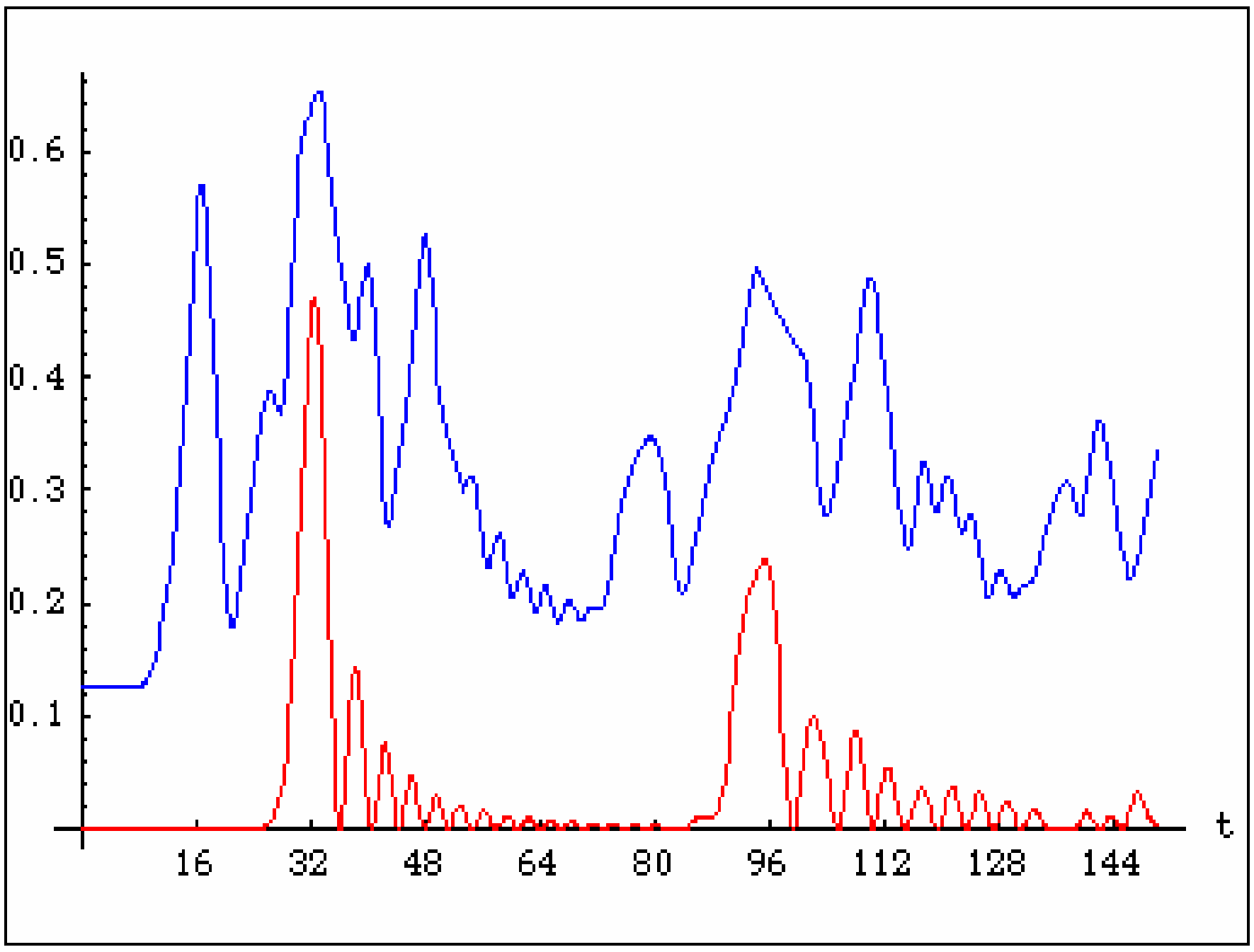

Figure 16
$\langle \psi(t)|P_a|\,\psi(t)\rangle$ *as a function of time. Figure* 15 *is reproduced for comparison.*

The behaviour of figure 8 shows the possibility of making full use of this advantage for large values of the number $2^K$ of applications of $AB$.

The mechanism through which the overlap probability $\Pr(t)$ can be made definitively close to $1/2$ is evident from $(3.32_b)$: one is taking, there, the average of $\left(\sin\left(\chi(\mu)\cdot\left(2\cdot n_j+1\right)\right)\right)^2$ over many periods, with respect to a probability measure $\frac{4\cdot j^2}{(\lambda\cdot t)^2} J_j(\lambda\cdot t)^2$ which is, for large values of $t$, fairly uniform over many periods, because of the wave packet spreading studied in [7].

In the context of figures 12 and 16, this simple argument is complicated by the details of the way in which $A$ and $B$ have been implemented in "local" terms.

Figure 16 corresponds to an implementation analogous to figures 8 and 9 in which the SWITCH primitives have been realized through spin raising and lowering operators $\sigma_\pm$.

A perfectly legitimate alternative consists in using, instead, projection operators. One can use, for instance, $\left(1+\sigma_3\right)/2$ instead of $\sigma_\mp$ in the upper branches of figures 8 and 9 and $\left(1-\sigma_3\right)/2$ instead of $\sigma_\pm$ in the lower branches.

Figure 17 shows the effect of this substitution. The improvement with respect to figure 16 is, of course, due to the fact that, now, the register qubits controlling the switches are not flipped during execution of A and B. The results of previous incomplete computations are, therefore, stored in a more efficient way.

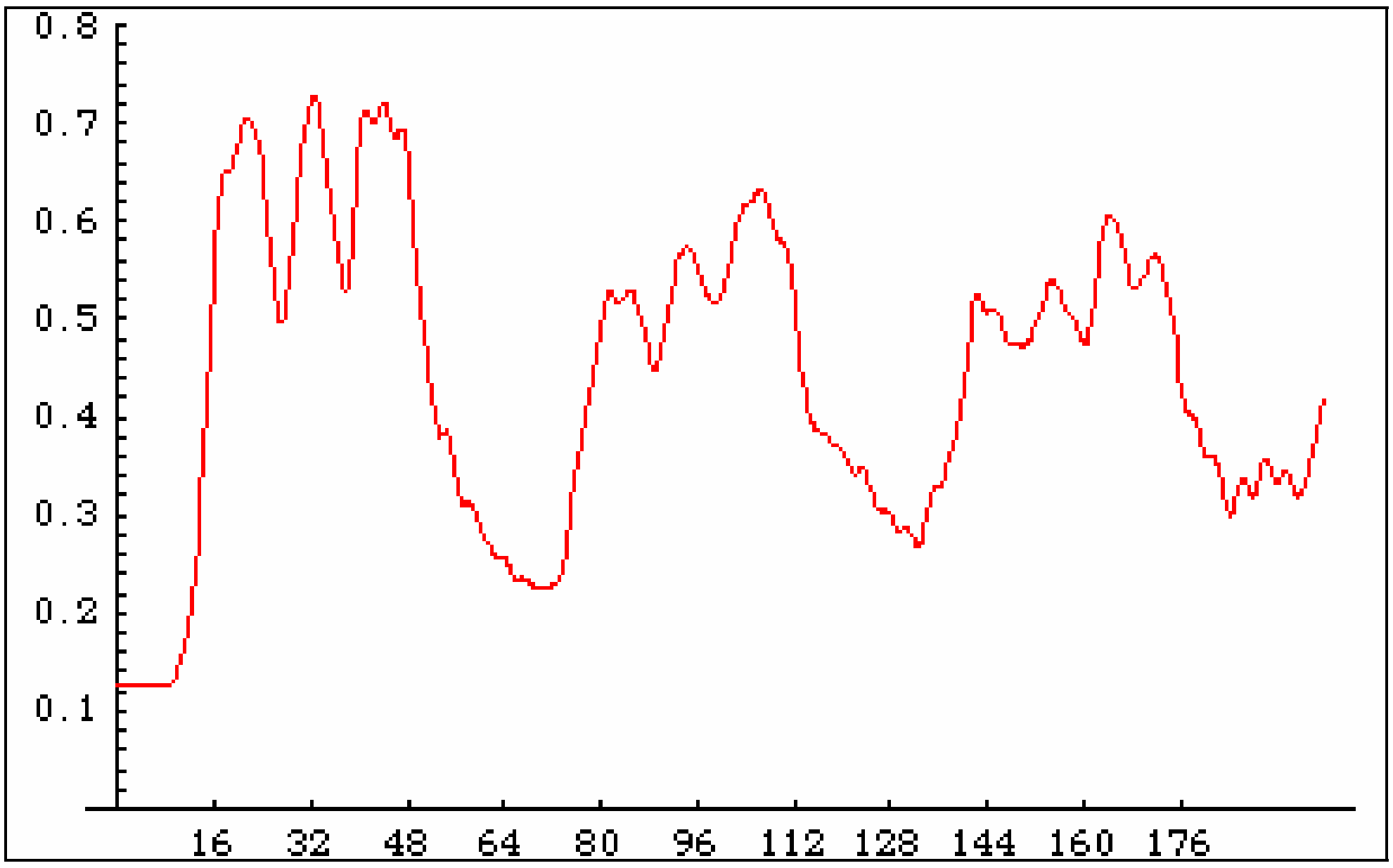

Figure 17

$\langle \psi(t)|P_a|\psi(t)\rangle$ *as a function of time, computed with the* SWITCH *primitives implemented as in figure* 18

Is it more convenient to implement the SWITCH primitives as in figure 9, with the controlling bit changing its state, or as in figure 18 ?

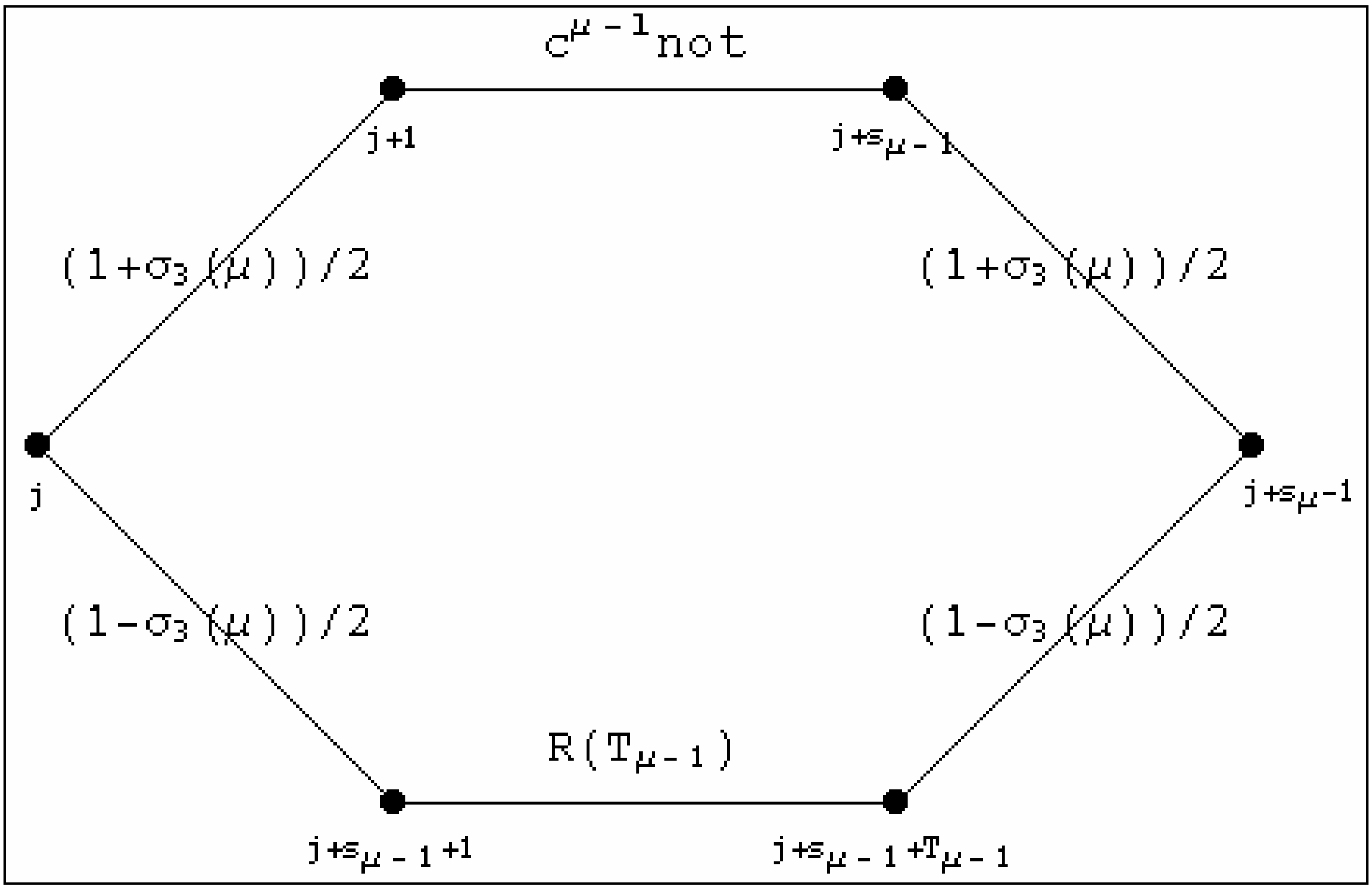

Figure 18

*As compared to figure 9, projectors, instead of creation and annihilation operators, are used here.*

If Grover's algorithm is a game in which Alice challenges Bob to find the word **a** hidden in the coupling constants of part A of figure 12, figure 9 would be a better strategy for Alice and figure 18 a better strategy for Bob.

But if, in figure 12, A is, say, a molecule waiting for an enzyme B to orient in some convenient way its chemical bonds, figure 18 would be, for both A and B, a better strategy.

And, we add as a concluding remark of this section, the pseudo-dissipative behaviour of figures 1, 3, 7, 16, 17 would help maintain a considerable level of overlap with the target orientation even after the optimal time $O\left(2^{\mu/2}\right)$ has elapsed.

## 6. Conclusions and outlook.

The SWITCH primitive can be implemented, at the quantum level, as a step of a continuous time quantum walk, namely as a term of the form $\sigma_-(j)\cdot\tau_+(f_1)\cdot\tau_-(i)+\sigma_+(j)\cdot\tau_+(f_2)\cdot\tau_-(i)$ in the register+cursor Hamiltonian.

The NOT primitive $\sigma_x(j)\cdot\tau_+(f)\cdot\tau_-(i)$ can be similarly viewed as a step of the quantum walk of the clocking excitation accompanied by the flipping of an assigned qubit of the register.

We have shown in section 4 that the CCNOT (or TOFFOLI) primitive and therefore every reversible computation can be described in terms of SWITCH and NOT.

The main body of this paper has been devoted to issues of clocking and synchronization of networks, such as the one in figure 12, made of the above two building blocks.

The term "synchronization" refers here to the systematic use we have made of delay lines in order to comply with the (essentially classical) requirement (or prejudice?) that all computational paths be of the same length: can one think of examples in which it is computationally advantageous to release this condition ?

On the explicit example of Grover's algorithm we have shown -see for instance figures 16 and 17 and the asymptotic behaviour of figure 3- that it <u>is</u> computationally advantageous to release the other classically "obvious" requirement that the output must be read from the register only after a measurement of the cursor has attributed a sharp value to the number of computational steps performed: the entanglement 2.19 or 3.15 between states of the clock and states of the register is indeed at the roots of the behaviour observed in the above figures.

As a final example of the "obvious" classical requirements that our model opens to criticism and scrutiny we call attention on the initial conditions $\left|\tau_3(1)=1,\tau_3(2)=-1\ldots,\tau_3(s)=-1\right\rangle$ in 2.13 and $\left|\tau_3(1)=1,\tau_3(2)=-1\ldots,\tau_3(s(K))=-1\right\rangle$ in 3.8: because of the conservation laws 2.15 and 3.10 they correspond to the "obvious" requirement of having just <u>one</u> clocking excitation. Could one release this requirement and allow, instead, the evolution 2.12 or 3.7 to take place out of the $\sum\frac{1+\tau_3(j)}{2}=1$ subspace? Stated more precisely: is there any example in which this is computationally advantageous ?

**Acknowledgements:**
It is a pleasure to thank Professor Ludwig Streit for his kind hospitality at CCM (Centro de Ciencias Matematicas, University of Madeira) during completion of this work.